\documentclass[conference]{IEEEtran}
\usepackage{cite}
\usepackage{amsmath,amssymb,amsfonts}
\usepackage{algorithmic}
\usepackage{graphicx}
\usepackage{textcomp}
\usepackage{xcolor}
\usepackage[hyphens]{url}
\usepackage{fancyhdr}
\usepackage{multirow}
\usepackage{tabularx}
\usepackage{booktabs}
\usepackage{colortbl}
\usepackage{threeparttable}
\usepackage{pifont}
\usepackage{soul}
\usepackage{bbding}
\usepackage{wasysym}
\usepackage[linesnumbered,lined,ruled,vlined]{algorithm2e}
\definecolor{mygray}{gray}{.92}
\definecolor{Common}{rgb}{0.9765, 0.9373, 0.7490}
\definecolor{RA}{RGB}{215, 248, 249}
\definecolor{RB}{RGB}{245, 232, 221}
\definecolor{RC}{RGB}{237, 164, 166}
\definecolor{RD}{RGB}{232, 232, 232}
\definecolor{RE}{RGB}{0, 255, 153}
\definecolor{RF}{RGB}{252, 140, 0}

\title{SOFA: A Compute-Memory Optimized Sparsity Accelerator via Cross-Stage 
Coordinated Tiling}

\author{\IEEEauthorblockN{Huizheng Wang\IEEEauthorrefmark{1}, Jiahao Fang\IEEEauthorrefmark{1}, Xinru Tang\IEEEauthorrefmark{1}, Zhiheng Yue\IEEEauthorrefmark{1}, Jinxi Li\IEEEauthorrefmark{1}, Yubin Qin\IEEEauthorrefmark{1}, Sihan Guan\IEEEauthorrefmark{1}, Qize Yang\IEEEauthorrefmark{1},\\ Yang Wang\IEEEauthorrefmark{1}, Chao Li\IEEEauthorrefmark{2}, Yang Hu\IEEEauthorrefmark{1}
and Shouyi Yin\IEEEauthorrefmark{1}
}
\IEEEauthorblockA{\IEEEauthorrefmark{1}\textit{School of Integrated Circuits, Tsinghua University, Beijing, China }\\
\IEEEauthorrefmark{2}\textit{School of Computer Science and Engineering, Shanghai Jiao Tong University, Shanghai, China}}
}

\begin{document}
\maketitle
\pagestyle{plain}


\begin{abstract}
Benefiting from the self-attention mechanism, Transformer models have attained impressive contextual comprehension capabilities for lengthy texts. 
The requirements of high-throughput inference arise as the large language models (LLMs) become increasingly prevalent, which calls for large-scale token parallel processing (LTPP). However, existing dynamic sparse accelerators struggle to effectively handle LTPP, as they solely focus on separate stage optimization, and with most efforts confined to computational enhancements. 
By re-examining the end-to-end flow of dynamic sparse acceleration, we pinpoint an ever-overlooked opportunity that the LTPP can exploit the intrinsic coordination among stages to avoid excessive memory access and redundant computation. Motivated by our observation, we present SOFA, a cross-stage compute-memory efficient algorithm-hardware co-design, which is tailored to tackle the challenges posed by LTPP of Transformer inference effectively. 
We first propose a novel leading zero computing paradigm, which predicts attention sparsity by using log-based add-only operations to avoid the significant overhead of prediction. Then, a distributed sorting and a sorted updating FlashAttention mechanism are proposed with cross-stage coordinated tiling principle, which enables fine-grained and lightweight coordination among stages, helping optimize memory access and latency. Further, we propose a SOFA accelerator to support these optimizations efficiently. Extensive experiments on 20 benchmarks show that SOFA achieves $9.5\times$ speed up and $71.5\times$ higher energy efficiency than Nvidia A100 GPU. Compared to 8 SOTA accelerators, SOFA achieves an average $15.8\times$ energy efficiency, $10.3\times$ area efficiency and $9.3\times$ speed up, respectively. 

\end{abstract}

\section{Introduction}\label{sec:introduction}
Remarkable success has been witnessed recently in the development of Transformer architecture~\cite{vaswani2017attention}, for both natural language processing (NLP)~\cite{brown2020language,devlin2018bert,lan2019albert,liu2019roberta,radford2018improving,radford2019language,raffel2020exploring,sanh2019distilbert,shoeybi2019megatron} and computer vision (CV) tasks~\cite{carion2020end,dosovitskiy2020image,li2022blip,liu2022swin,liu2021swin,radford2021learning,rombach2022high,zhai2022scaling,zhu2020deformable}. 
The impressive capabilities of Transformers greatly stems from their \emph{self-attention} module, which excels at extracting global context information~\cite{wu2019lite}. Typically, self-attention modules take three matrices as their inputs: namely, $\mathbf{Q}$ (query), $\mathbf{K}$ (key) and $\mathbf{V}$ (value). First, an attention matrix $\mathbf{A}$$\in\mathbb{R}^{S\times S}$ is obtained by multiplying $\mathbf{Q}$ and $\mathbf{K}$, where $S$ is sequence length. Next, $\mathbf{A}$ goes through the softmax function for normalization, then is multiplied by $\mathbf{V}$ for the final output. 




Large language models (LLMs) have driven the transformer architecture to unprecedented levels of complexity and capability, particularly in handling extended sequence lengths\cite{ladhak-wiki-2020}. This evolution places heightened demands on inference capabilities and throughput\cite{minaee2024large}, critically impacting the performance of key transformer components: the attention module, feed-forward network (FFN) module, and the query-key-value (QKV) computations. 

\begin{figure}[t]
\centering
\includegraphics[width=0.9\linewidth]{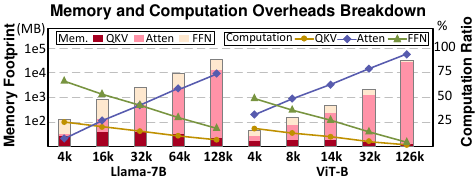}\vspace{-2mm}
\caption{Transformer memory and computation breakdown for long sequence.}
\label{fig:Intro}\vspace{-4mm}
\end{figure}

Traditionally, in Transformers designed for smaller sequence lengths($\le$\textbf{2k}), the FFN module typically presented the main bottleneck due to its dense computational requirements. However, with recent advancements in processing long text, where sequence lengths can exceed 128,000 characters\cite{li2024long,wu2024loongserve,liu20242}, the performance bottleneck is shifting from the FFN to the attention module. Our detailed profiling indicates that as sequence lengths surpass 32,000 characters, the attention module becomes the dominant factor affecting inference time, as shown in Fig.\ref{fig:Intro}. This shift is primarily because the complexity of the attention mechanism scales quadratically with sequence length, making it increasingly challenging to manage as sequences extend. 

\emph{Dynamic sparsity (DS) acceleration}~\cite{ham20203,ham2021elsa,lu2021sanger,qu2022dota,yang2022dtatrans,qin2023fact,wang2021spatten,zhou2022energon} have emerged as a promising solution to mitigate the latency issue of self-attention. The key idea is to predict vital Q-K pairs at runtime and calculate attention based on these vital pairs to reduce the inference latency. Typically, it consists of three stages. A \emph{pre-compute stage}  firstly estimates the matrix $\mathbf{A}$ (denoted as $\hat{\mathbf{A}}$). Then, a \emph{top-k stage} picks the vital Q-K pairs. In the subsequent \emph{formal computing stage}, self-attention is calculated only based on the vital pairs. 

The need for \textbf{\textit{high parallelism of dynamic sparsity token processing in the context of LLM inference}} is increasing, especially during the prefill stage. In this stage, entire contexts are processed simultaneously, favoring high token parallelism to enhance efficiency. This scenario is especially meaningful as modern LLM inference often employs separate deployments for the prefill and decode stages\cite{zhong2024distserve,patel2023splitwise}. Moreover, the advent of speculative inference\cite{zhao2024ouroboros} can transform decode operations into prefill tasks, further emphasizing the need for efficient large-scale token processing parallelism (LTPP).

\begin{figure*}[t]
\centering
\includegraphics[width=0.95\linewidth]{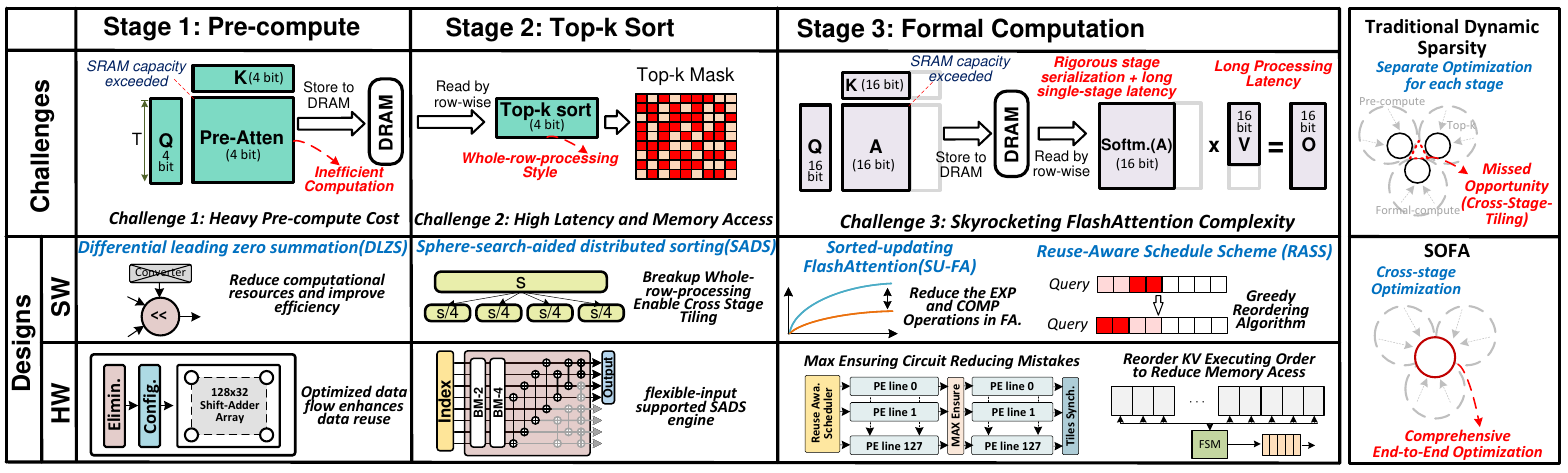}\vspace{-2mm}
\caption{Dynamic sparsity challenges for LTPP and SOFA's software and hardware co-design.}
\label{fig:workflow}
\end{figure*}

However, supporting dynamic sparsity with large-scale token parallel processing would present prohibitive overheads, as shown in Fig.~\ref{fig:workflow}. This is because, firstly, current dynamic sparsity acceleration solutions lack efficient prediction schemes to reduce computation complexity. Though calculating self-attention based on vital Q-K pairs can be beneficial in reducing compute and memory consumption, the newly introduced \emph{pre-compute} and \emph{top-k} stages consume non-trivial computational and memory resources when large amounts of tokens are processed, which can even offset the benefits brought by sparsity acceleration methods in some cases. Our characterization depicts that even with 4-bit during the \emph{prediction stage} and 16-bit during the \emph{formal stage}, the power overhead of prediction is already $1.4\times$ that of formal computing when top-$k$ equals $20\%$. Unfortunately, the overhead in prediction will further rise sharply with increased parallelism.

\begin{figure}[t]
\centering
\includegraphics[width=0.95\linewidth]{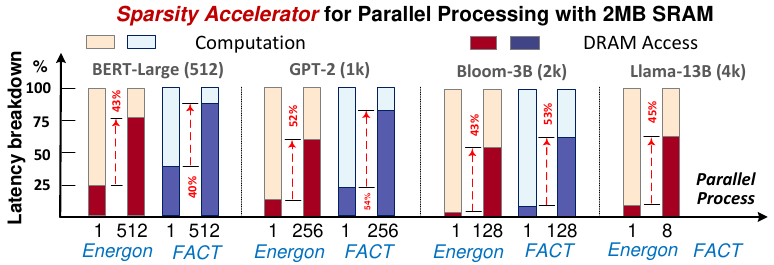}\vspace{-2mm}
\caption{MAT for SOTA dynamic sparsity accelerators (FACT \cite{qin2023fact}, Energon \cite{zhou2022energon}) with diverse parallelisms.}
\label{fig:SOFA}
\end{figure}

Secondly, the processing stages in current dynamic sparsity acceleration are not designed to be partitionable, and miss the opportunity to support fine-grained pipelining, which would enable more efficient processing. The top-k sorting must be based on the readiness of the whole row of Pre-Atten matrix. In LTPP scenarios, the increased delay in processing each stage accumulates continuously, ultimately resulting in a significant increase in end-to-end latency. This "whole-row-processing" style also increases the amount of intermediate data, resulting in a substantial rise in DRAM access requirements. Fig.~\ref{fig:SOFA} shows the memory access time (MAT) of two SOTA accelerators when scaled to process multiple tokens. The increase in parallelism leads to a sharp rise in off-chip memory access and surging MAT. On average, the MAT ratio rises to $72\%$, overshadowing computation time and becoming the primary bottleneck.

Thirdly, current dynamic sparsity acceleration solutions do not exploit cross-stage coordination, missing the opportunity to reduce the computation complexity of later stages by leveraging guidance extracted from former stages. Although FlashAttention2 (FA-2) already provides a tiling scheme for softmax to reduce memory access overhead, \textbf{\textit{the decreased memory access comes with surging computations}}. This occurs because repeated exponentiation and comparison operations are necessary to refresh the MAX among tiles, ensuring the correctness of the global MAX value. We observe an opportunity to guide FA-2 computation with top-$k$ information. These limitations highlight the need for more advanced strategies to manage dynamic sparsity with LTPP effectively.

\emph{\textbf{\textit{Our Insights: }}}
Motivated by the challenges, we observe an opportunity that breaks down the computation, memory, and latency overheads in each stage by adopting a cross-stage coordinated tiling strategy, thus a stage is decomposed into fine-grained sub-stages. Therefore the process in the following stages doesn't have to wait for the finish of processes in the last stage. The coordination among stages becomes more swift and excessive DRAM memory access could be saved. Notably, it is non-trivial to achieve this goal as we need to figure out effective methods to partition top-k module and efficiently forward the information to formal stages. 

We propose an algorithm-hardware co-design for attention optimizations, named SOFA. 
It features three key designs that correlate to three challenges, as depicted in Fig.~\ref{fig:SOFA_workflow}. 
First, the computation overhead in pre-compute stage is alleviated via a multiplier-free \emph{differential leading zero summation (DLZS)} paradigm, which helps reduce the sparsity prediction overhead of each tile. Second, 
we propose a \emph{sphere-search-aided distributed sorting (SADS)}, which distributes a long segment into sub-segments to execute individual tiled sorting, while effectively reducing total comparisons. Third, we propose a \emph{sorted-updating FlashAttention (SU-FA)}. It skillfully decouples the \texttt{softmax} row-dependence to enable the formal computing stage tiling, while leveraging cross-stage sorting information to reduce computation. In summary, DLZS and SADS together serve as a low-complexity prediction (LP) mechanism to reduce prediction overhead. SADS collaborates with SU-FA, employing fine-grained tiling for sparse acceleration, to optimize memory access and processing latency.

We propose a dedicated accelerator to support the proposed mechanism effectively. Compared to naive implementation, which only has a limited $19.6\times$ energy saving over Nvidia A100 GPU, SOFA accelerator improves its performance with four novel algorithm-hardware co-designs. 

Evaluated on $20$ benchmarks, SOFA achieves an average energy efficiency of $7183$ GOPS/W, which is $71.5\times$ and average $15.8\times$ higher than Nvidia A100 GPU and 8 SOTA accelerators, respectively. Overall, SOFA outperforms GPU A100 $9.5\times$ and TPU $11.1\times$, respectively. We also conduct comprehensive ablation on GPU to quantify the performance benefits brought by our software mechanism and various hardware components. Evaluations on GPU/TPU show, SOFA's software optimization provides $3.16\times/2.8\times$ speedup, while hardware acceleration enables a $3.03\times/3.9\times$ speedup. 

\section{Background and Motivation}\label{sec:background}

\subsection{Preliminaries for Transformer}\label{subsec:computation_transformer layer}
\begin{figure}[t]
\centering
\includegraphics[width=0.99\linewidth]{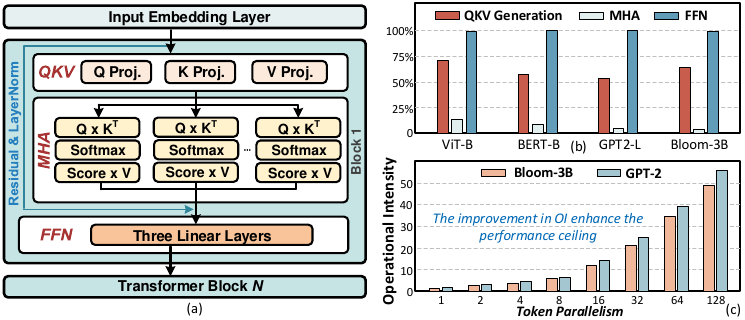}\vspace{-2mm}
\caption{Basic components of a Transformer model and operation intensity.}
\label{fig:Transformer}
\end{figure}

Fig.~\ref{fig:Transformer}(a) shows a typical Transformer model: an input sequence containing $S$ tokens is transformed into an embedding matrix $\mathbf{X}\in \mathbb{R}^{S\times H}$, projected to $\mathbf{Q}$, $\mathbf{K}$ and $\mathbf{V}$ spaces, split into $A$ chunks $\mathbb{R}^{S\times H/A}$, and processed by multi-head attention (MHA) to generate an attention matrix. The attention matrix, after softmax and multiplication with $\mathbf{V}$, resulting in a matrix $\mathbf{O}\in\mathbb{R}^{S\times (H/A)}$. Outputs from all heads are concatenated, projected by $\mathbf{W}_O\in\mathbb{R}^{S\times H}$, and passed through the FFN with two fully connected layers to generate final outputs. 


\textbf{Computation Properties Analysis.} We analyze the operation intensity (OI)~\cite{williams2009roofline} for the three parts of a Transformer layer. As shown in Fig.~\ref{fig:Transformer}(b), MHA exhibits notably lower OI, averaging $15\%$ of the FFN. This means MHA requires more data movement for the same computation FLOPs, due to element-wise operations. Fig.~\ref{fig:Transformer}(c) further illustrates the relationship between the OI of MHA and the token processing parallelism. We can figure increasing parallelism effectively boosts OI, thus theoretically reducing the demand for data movement under equivalent computational power and PE utilization. This gain is attributed to increased data reuse.

\subsection{FlashAttention (FA)}\label{subsec:flashattention}
\begin{figure}[t]
\centering
\includegraphics[width=0.95\linewidth]{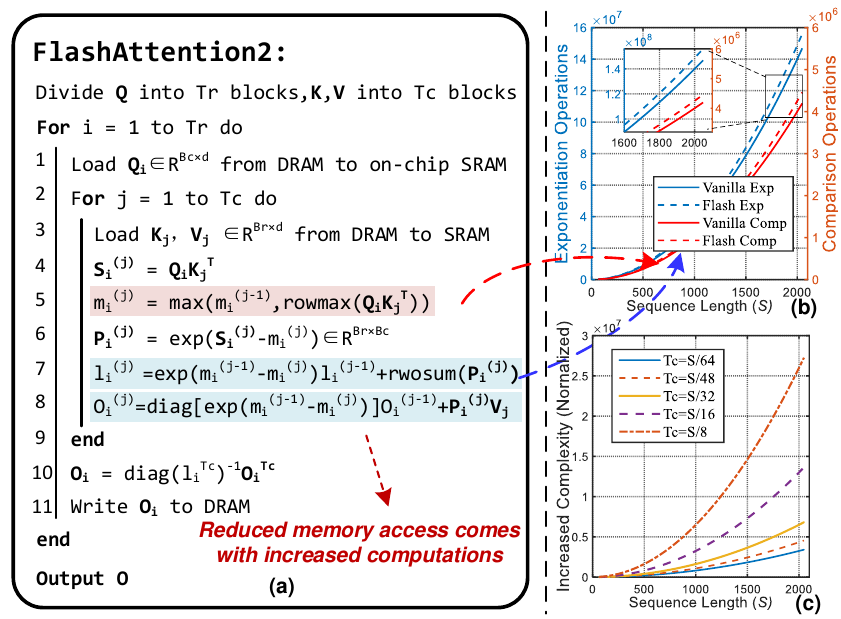}\vspace{-1mm}
\caption{Process of FlashAttention-2 and its computation overhead.}
\label{fig:Flash_Cost}
\end{figure}

To reduce data movement of attention, Tri Dao \emph{et. al} proposed FlashAttention (FA)\cite{dao2022flashattention} and improved version FA-2\cite{dao2023flashattention}, both of which successfully minimized memory access but greatly increased computational cost. Fig.~\ref{fig:Flash_Cost}(a) outlines the procedure of FA-2 and Fig.~\ref{fig:Flash_Cost}(b) compares its exponential operations and comparison complexity with vanilla implementation regarding $S$. Here we assume the number of tiles $T_c$=$S/16$, i.e., tiling size $B_c$=$16$. We employ the arithmetic complexity model~\cite{brent2010modern} to normalize the complexity for different operations. As $S$ increases, FA-2 exhibits a notable increase in exponential and comparison operations compared to the vanilla scheme. When $S$=$2048$, it demands $9\times10^6$ more exponential calculations and $3\times10^5$ more comparisons than the vanilla implementation. Fig.~\ref{fig:Flash_Cost}(c) compares the increased computational load after summing all calculations. The computational complexity of FA-2 soars with the growth of $S$, and the increased magnitude correlates with $T_c$. The larger $T_c$ leads to a faster increase, due to the repeated calculations among $T_c$ blocks, as shown in lines $5$-$8$ of Fig.~\ref{fig:Flash_Cost}(a).

\subsection{Sparsity in Attention}\label{subsec:Sparse_Attention} 

Typically, as shown in Fig.\ref{fig:Transformer} (a), the results (a.k.a scores) of $\mathbf{Q}\times \mathbf{K}^T$ are then processed by a softmax operator. Due to the \emph{softmax}'s approximation to the \emph{argmax} operator, most smaller score values become extremely close to zero after passing the \emph{softmax}. Therefore, they usually impose a negligible impact on the final results and can be reasonably removed. The difference between the attention sparsity and DNN/Transformer model sparsity is that attention sparsity is entirely determined by the input data and requires dynamic judgment at runtime, whereas model sparsity is based on static weight sparsity, which can be optimized through quantization or structured pruning.    

To accelerate \emph{self-attention}, emerging dynamic sparsity accelerations~\cite{ham20203,ham2021elsa,wang2021spatten,lu2021sanger,zhou2022energon,qin2023fact,qu2022dota} offer a promising solution. Their key idea is to predict key Q-K pairs at runtime and calculate attention for selected pairs. Typically, their workflow proceeds as Fig.~\ref{fig:SOFA_workflow}. First, a low-precision computational paradigm is employed to predict the attention (\emph{Pre-compute stage}); Next, vital Q-K pairs are filtered out from each row to generate a mask (\emph{Top-k sorting stage}). Finally, based on the mask, the scheduler initiates the \emph{Formal Computing Stage}, typically with higher precision.

\subsection{Analysis for Large-scale Token Parallel Processing (LTPP)}\label{subsec:Token_Parallelism_Analysis}
 Despite promising adaptability, dynamic sparsity incurs additional overhead (\emph{Pre-compute} and \emph{Top-$k$ stages}) during inference. As a result, previous works~\cite{ham20203,wang2021spatten,zhou2022energon,ham2021elsa,qin2023fact} were limited to processing queries with low parallelism, to minimize the memory and computation overhead. However, as modern LLMs demand significantly longer context than before (GPT$4$ $32$k \cite{achiam2023gpt}, LongLLaMa $256$k \cite{tworkowski2024focused}), the rapid processing of long context becomes increasingly crucial~\cite{kachris2024survey}. This highlights the necessity for accelerators with LTPP capabilities. However, the current dynamic sparsity attention workflow poses three challenges for LTPP. Illustrated in Fig. \ref{fig:workflow}: 

1) Supposing processing $T$ tokens in parallel, the pre-compute and sorting complexity rises to $\mathcal{O}(TSH)$ and $\mathcal{O}(TSSk)$, respectively. Taking Llama-13B ($T$=$512$,$k$=$0.25$) as an example, the required numbers of comparisons and multiplication would be over $10^{11}$ and $10^8$, respectively. In this case, prediction requires performing over $2^{11}$ MACs and $2^{10}$ comparisons, accounting for more than $57\%$ of the total execution latency. Such prohibitive overhead will negate the improvements brought by sparsity.


2) As \emph{top-k} sorting and \emph{softmax} is applied row-wise, matrices \textbf{Pre-Atten} and \textbf{A} must be stored to DRAM first and then loaded by row blocks, thus leading to massive DRAM access. Such extensive memory access would lead to inefficient inference. In $45$\,nm CMOS technology, the energy cost of a DRAM access ($5$ to $20$ pJ/bit) is two orders of magnitude higher than that of internal cache access ($0.1$ pJ/bit)\cite{horowitz20141}, while its bandwidth (DDR4 $25.6$GB/s) is also orders-of-magnitude lower than the SRAM ($19$TB/s)~\cite{dao2022flashattention}. A coarse scheme is to enlarge the on-chip SRAM capacity but this would lead to area inefficiency. Taking ($T$=$512$, $S$=$2048$) for instance, it directly necessitates $5$MB SRAM, leading to $5.47$ mm$^2$ footprint under TSMC 28nm technology, which is $7.4\times$, $8.9\times$ of the overall area of SpAtten\cite{wang2021spatten} and ELSA\cite{ham2021elsa}, respectively.


c) FlashAttention2 (FA-2) employs a tiling scheme for the softmax operation to keep the working set of data in the faster on-chip memory, thus successfully reducing memory access overhead. However, the benefits come with soaring computation costs, making it unsuitable for dynamic sparsity scenarios in LTPP. As an example, when the tile size is $B_c=4$ for a sequence length $SL$=1024, FA-2 must frequently compute and compare values across these tiles to ensure correct global results. This leads to a computational load approximately $1.5\times$ higher than that of a regular implementation without tiling.



\textbf{\emph{We argue that the main bottleneck in extending existing dynamic sparsity methodology towards LTPP lies in information decoupling among stages, thus missing the cross-stage-tiling opportunity.}} Table~\ref{tab:works_comparision} offers an overview of the effectiveness of existing approaches in optimizing Transformer components. 
Works\cite{ham20203,ham2021elsa,lu2021sanger,qu2022dota} focus on reducing pre-computation overhead, such as ELAS\cite{ham2021elsa} using Binary Hash, A3\cite{ham20203} employing Greedy search and Dota using low-rank transformation\cite{qu2022dota}. However, these methods still cannot address the row dependency of key operators, like top-$k$ and \emph{softmax}, thus resulting in significant memory access overhead under LTPP. Further, SpAtten\cite{wang2021spatten} and DTATrans\cite{yang2022dtatrans} involve sorting the cumulative distribution probabilities of tokens, introducing substantial sorting complexity and latency in LTPP scenarios.
While SpAtten\cite{wang2021spatten} and Energon\cite{zhou2022energon} realize challenges with extensive memory access, their sparsity strategies fail to handle the severe memory access overhead with the LTPP scenario. In summary, existing works are all limited on individual-stage optimization, thereby overlooking opportunities for cross-stage joint optimizations, making them inadequate for supporting LTPP. This motivates us to propose a cross-stage compute-memory efficient accelerator design, targeting the LTPP scenario. 


\begin{table}[t]
\renewcommand{\arraystretch}{0.9}
\caption{Summary for SOTA Transformer Accelerators.}\vspace{-3mm}
\begin{center}
\begin{tabular}{l||m{0.5cm}<{\centering}|m{1.2cm}<{\centering}|m{0.5cm}<{\centering}|m{1.2cm}<{\centering}|m{0.7cm}<{\centering}}
\specialrule{0.12em}{0.5pt}{0.5pt}
\multirow{3}{*}{\!\!\!\textbf{Accelerator}} & \multicolumn{5}{c}{\textbf{Optimization}}\\
\cline{2-6}
&   \multicolumn{2}{c|}{Compute} & \multicolumn{2}{c|}{Memory} & Cross\\
\cline{2-5}
& \multirow{1}{*}{\!\!QKV}  & Attention & QKV   &  \multirow{1}{*}{Attention} & Stage\\
\hline
\rowcolor{mygray}\!\!\!$\mathbf{A}^3$\!\cite{ham20203} & $\times$ & \checked  & $\times$ & $\times$ & $\times$ \\
\!\!\!\textbf{ELSA}\cite{ham2021elsa} & $\times$ & \checked  & $\times$ & $\times$ & $\times$\\
\rowcolor{mygray}\!\!\!\textbf{Sanger}\cite{lu2021sanger} & $\times$ & \checked  & $\times$ & $\times$ &$\times$ \\
\!\!\!\textbf{DOTA}\cite{qu2022dota} & $\times$ & \checked  & $\times$ & $\times$ & $\times$\\
\rowcolor{mygray}\!\!\!\textbf{Energon}\cite{zhou2022energon} & $\times$ & \checked  & $\times$ & Low & $\times$\\
\!\!\!\textbf{DTATrans}\cite{yang2022dtatrans}\!\!\! &  $\times$ & \checked  & $\times$ & $\times$ & $\times$\\
\rowcolor{mygray}\!\!\!\textbf{SpAtten}\cite{wang2021spatten} & \checked & \checked  & $\times$ & Low & $\times$\\
\!\!\!\textbf{FACT}\cite{qin2023fact} & \checked & \checked  & $\times$ & $\times$ & $\times$ \\
\rowcolor{mygray}\!\!\!\textbf{SOFA} & {\fontsize{6.5}{5} \CheckmarkBold }& {\fontsize{6.5}{5} \CheckmarkBold }  & {\fontsize{6.5}{5} \CheckmarkBold } & {\fontsize{6.5}{5} \CheckmarkBold } & {\fontsize{6.5}{5} \CheckmarkBold }\\
\specialrule{0.12em}{0.5pt}{0.6pt}
\end{tabular}
\end{center}
\label{tab:works_comparision}\vspace{-2mm}
\end{table}

\section{Algorithm Optimizations of SOFA}\label{sec:SOFA_algorithm}

\begin{figure}[t]
\centering
\includegraphics[width=0.98\linewidth]{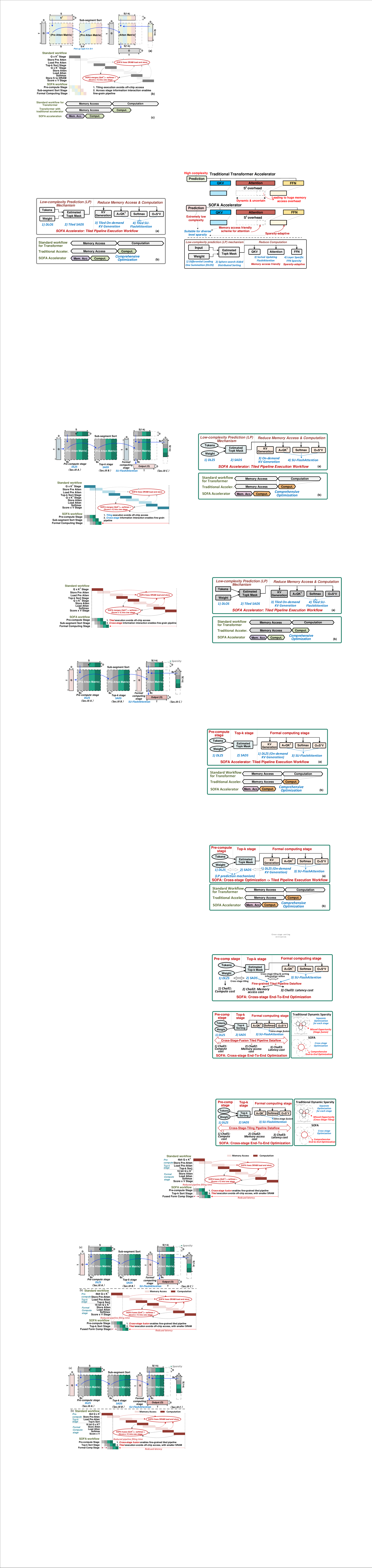}\vspace{-2mm}
\caption{(a) High-level diagram of the SOFA algorithm optimizations. (b) Tile-based pipelined dataflow (SOFA) vs. standard dataflow.}
\label{fig:SOFA_workflow}\vspace{-2mm}
\end{figure}

Fig.~\ref{fig:SOFA_workflow} (a) presents an overview of the SOFA algorithm optimizations. First, at the \emph{pre-compute} stage, we propose DLZS, a log-domain computing paradigm named DLZS to predict $\mathbf{\hat{A}}$. Then, exploiting \emph{DCE}, we introduce SADS, to partition a long sequence into several sub-segments for independent tiled sorting. Next, leveraging the sorting information, a memory-compute efficient attention-computing mechanism (SU-FA) is designed. 
The SADS and SU-FA enable SOFA to execute a cross-stage tiling pipeline dataflow. Compared to the vanilla workflow in Fig.~\ref{fig:SOFA_workflow}(b), the tiling execution makes SOFA require minimal SRAM for storing intermediate results without extra memory access, while the fine-grained pipelined dataflow can reduce inference latency. 

\begin{figure}[t]
\centering\includegraphics[width=0.99\linewidth]{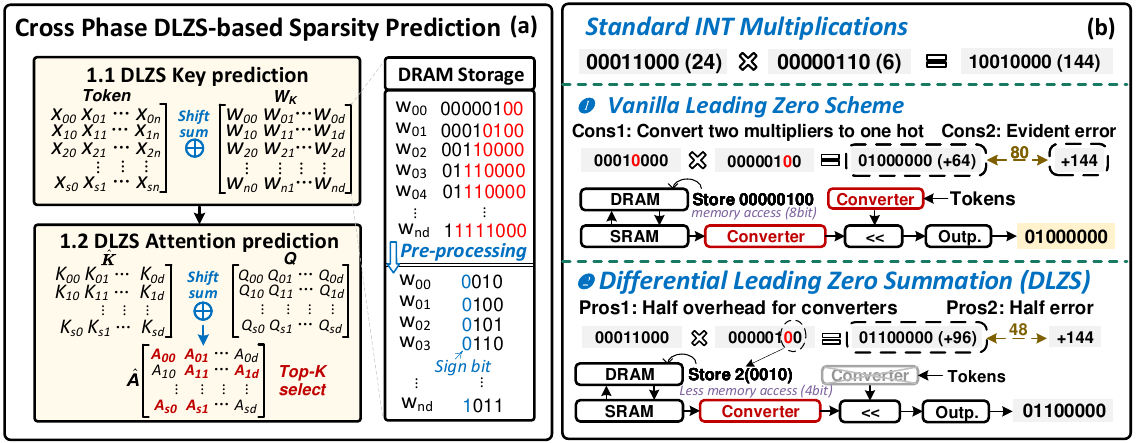}\vspace{-2mm}
\caption{(a) Cross-phase DLZS sparsity prediction. (b) Comparisons between DLZS and vanilla scheme.}
\label{fig:DLOS}
\end{figure}


\subsection{Cross-Phase DLZS Sparsity Prediction}\label{subsection:DLZS}

Traditional \emph{dynamic sparsity} entails predicting significant Q-K pairs, then utilizing these important Ks and Vs to execute computations. However, blindly generating unnecessary KV leads to wastage in computation and memory access. To this end, SOFA employs an \emph{on-demand} computation strategy for KV. As shown in Fig.~\ref{fig:DLOS}(a), \emph{On-demand} means: only the required Ks and Vs are generated ($\mathbf{K}_i$=$\mathbf{x}_i\mathbf{W}_k$,$\mathbf{V}$=$\mathbf{x}_i\mathbf{W}_v$), while trivial ones are not computed from the beginning. However, this requires the \emph{pre-compute stage} first to estimate the $\hat{\mathbf{K}}$, then utilize it with $\mathbf{Q}$ to predict $\hat{\mathbf{A}}$. Unfortunately, even utilizing low-precision matrix multiplication (e.g. half-precision with MSBs only) results in considerable power consumption. Therefore, a power and memory-efficient prediction is imperative.    

We propose a log-domain multiplication-free strategy, named differential leading zero summation (DLZS). \emph{Differential} means: For multiplication, it only transforms one operand into the logarithmic domain using the leading zero encoder (LZE), to obtain its leading zero (LZ). Then, based on the LZ, it substitutes
the costly multiplication with low-power shift operations on the other operand. Specifically, an INT-type number $x$ can be mathematically expressed as Eq.~\eqref{eq:LoD1}, where $W$ stands for the bit-width, $M$ represents the mantissa lying $[0,\,1]$, and $LZ$ denotes the leading-zero count of $x$. Accordingly, the corresponding multiplication is derived as Eq.~\eqref{eq:LoD3} and approximated as Eq.~\eqref{eq:LoD4}. Since the bit width $W$ is fixed for certain operands, we can directly operate $LZ_y$ on $x$, to estimate the magnitude for the product of two numbers. Therefore, incorporating shifting and the sign bit, the results of multiplication can be predicted.   
\begin{subequations}
\begin{align}
&x=Sign\times M\times2^{W-LZ},\label{eq:LoD1} \\
&x\cdot y=\texttt{XOR}\left(S_x,S_y\right)M_x\cdot2^{(W-{\scriptsize\vspace{-3mm}LO_x)}} M_y\cdot2^{(W-{\scriptsize\vspace{-3mm}LO_y)}}\label{eq:LoD3}\\
&~~~~~~\approx\texttt{XOR}\left(S_x,S_y\right)M_x\cdot2^{(W-{\scriptsize\vspace{-3mm}LO_x+W-LO_y)}}\label{eq:LoD4}
\end{align}
\end{subequations}


Its workflow is depicted in Fig.~\ref{fig:DLOS}(b). As the weights are pre-known and fixed during inference, we pre-convert the $\mathbf{W}_k$ into LZ format and store it. Then, in the \emph{Key prediction phase} (1.1), no LZE is required. In the subsequent \emph{Attention prediction phase} (1.2), to mitigate error accumulation, we convert $\mathbf{Q}$ into the log domain instead of $\hat{\mathbf{K}}$, then perform shifting and sum operations. Compared to the vanilla leading zero strategy (Fig.~\ref{fig:DLOS}(c)), the proposed DLZS exhibits three Pros: a) Lower converter overhead; b) Higher accuracy; c) Less memory access.    

\subsection{Sphere-search Aided Distributed Sorting (SADS)}\label{subsec:SADS}

\begin{figure}[t]
\centering
\includegraphics[width=0.85\linewidth]{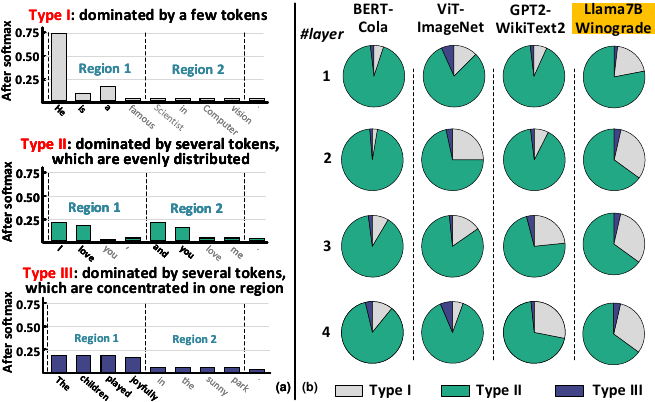}\vspace{-2mm}
\caption{(a) Three types of attention data distribution. (b) Corresponding proportions in diverse Transformer models.}
\label{fig:statistical_result}
\end{figure}

As \texttt{softmax} approximates the \texttt{argmax} operation, its results primarily depend on dominant tokens when multiple tokens with prominent amplitudes appear, as denoted in Type-I of Fig. \ref{fig:statistical_result}(a). Alternatively, there are two potential scenarios for element distribution: a uniform distribution, exemplified by Type-II, and a concentration of slightly larger elements in a specific region, depicted as Type-III. To ascertain their practical distributions in Transformer inference, we conducted a token analysis for BERT/L~\cite{devlin2018bert}, ViT/B~\cite{dosovitskiy2020image}, GPT-2~\cite{radford2019language}, Llama7B with $4096$ rows. The statistical results in Fig.~\ref{fig:statistical_result}(b) reveal that the Type-II distribution predominates across all four models, accounting for over $76\%$ on average. Type-I occurrence is more frequent in ViT, GPT-$2$ and Llama, with an average rate of $25\%$, which may be attributed to image local similarity and the self-autoregressive token generation, respectively. By contrast, the occurrence probability of Type-III is notably low in all models, even approaching nearly $0$ in GPT-$2$ and Llama. This is primarily attributed to the extended context, which diminishes the likelihood of a concentration for higher magnitude tokens in a specific region. 


Combined, Type-I and Type-II together make up over $95\%$ of the total distribution, effectively representing the overall data distribution characteristics of attention. The larger values within each region of these two data types can aptly represent the overall larger values. We term this characteristic as the `\emph{Distributed Cluster Effect (DCE)}'. \emph{Distributed} implies that a long segment can be divided into several shorter sub-segments, while \emph{Cluster} indicates that each sub-segment contains its primary information. Therefore, sorting based on well-segmented partitions is expected to have a negligible impact on holistic performance. 

\begin{figure}[t]
\centering
\includegraphics[width=0.95\linewidth]{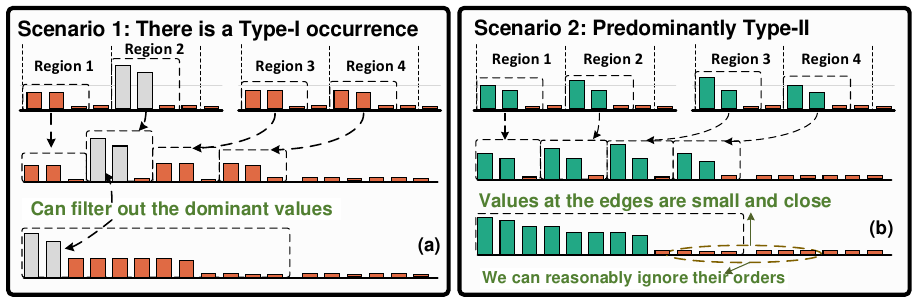}\vspace{-2mm}
\caption{(a) Scenario 1: Type-I occurs. (b) Scenario 2: Type-II dominates.}
\label{fig:SADS}
\end{figure}

To this end, we propose the SADS sorting, which exploits the \emph{DCE} to reduce complexity in a tiled manner. Initially, one row of the attention matrix is divided into $n$ sub-segment (assuming $n$=$4$). Next, each sub-segment pick up the top-$(k/n)$, i.e., top-$(k/4)$ values, from its own data. Following this, for each sorted set, the largest $k/4$ elements are collected into $\mathbf{FC}$ set, which represents the indices of vital KVs. This set is used to guide the subsequent \emph{Formal Computing Stage}. 


Figs.~\ref{fig:SADS}\,(a)-(b) exemplify why SADS can maintain accuracy with reduced complexity. For  Scenario $1$, where Type-I distribution occurs, SADS is certain to capture the dominant values, irrespective of which sub-segment they fall into. For Scenario $2$, where the majority of the distribution is Type-II, SADS can effectively select all relatively larger values that dominate in the complete row. Given that the values falling on the edges of the top-$k$ are typically smaller, we can reasonably relax the sorting requirements for them. Furthermore, the specific number of sub-segments (e.g. tiling) of each layer is obtained by the DSE in Section~\ref{sec:DSE}.



\subsection{Sorted-Updating FlashAttention (SU-FA)}\label{subsec:SU-FlashAttention}
The attention is the primary bottleneck in scaling to LTPP, as its memory complexity increases quadratically with the sequence length.
To tackle this issue, we propose an attention acceleration mechanism called SU-FA, which is computationally and memory efficient, by leveraging specific sorting information generated from the \emph{top-k stage}. It also enables cross-stage tiling for the formal-compute stage. Traditionally, addressing overflow in hardware \texttt{softmax} implementation requires identifying the Max value in each row. This necessitates continual comparisons in classical FA~\cite{dao2022flashattention,dao2023flashattention} to refresh the Max value across diverse blocks, which however, results in skyrocketing computational cost as revealed in Fig.~\ref{fig:Flash_Cost}.


\begin{figure}[t]
\centering
\includegraphics[width=0.92\linewidth]{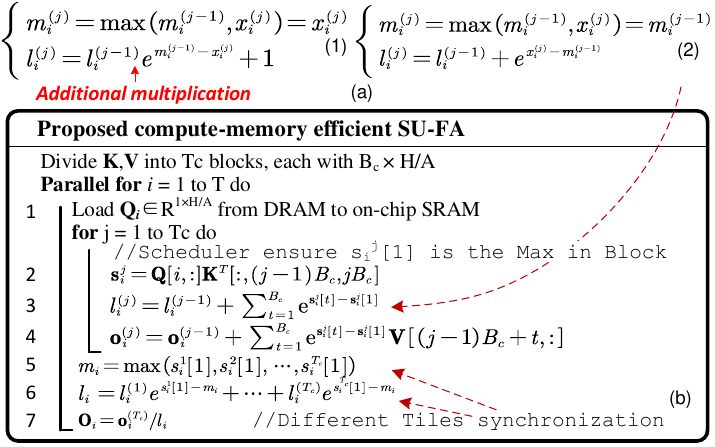}\vspace{-2mm}
\caption{(a) Formulas for diverse updating orders. (b) Procedure of SU-FA.}
\label{fig:SU-FA}
\end{figure}

The indices of the top-k values provided by \emph{top-k stage} allow us to get the potential index of the Max value. A direct but coarse approach is to calculate the Max value based on the potential index and then send it into the FA for computation. However, there are two critical problems: 1) The index of the Max is not guaranteed to be accurate due to the approximation of DLSZ, leading to overflow; 2) Calculating the Max separately introduces additional computation and power overhead. To this end, we propose a novel sorted-updating FA. Instead of computing for the Max separately, SU-FA executes either ascend or descend updates during the computation process. Descend updating means first computing Fig. \ref{fig:Flash_Cost}(a) line 5 from the index of Max, followed by the index of the 2nd large value, until the $k$-th value. Ascend updating proceeds in the opposite order. Although at first glance, both of these approaches can effectively eliminate the $\texttt{max}$ comparison (Fig. \ref{fig:Flash_Cost} (a) line5), we found that the benefits vary significantly with different updating orders. Specifically, when executing ascend updating, the line $5$-$7$ can be rewritten as Eq.\,(1) in Fig.~\ref{fig:SU-FA}(a), where we denote $\mathbf{S}_i^{j}$ as $x_i^{(j)}$ for clarity. Though $m_i^{(j)}$ equals to $x_i^{(j)}$ constantly, it is noteworthy the calculating for $l_i^{(j)}$ still acquires one exponentiation (Exp), one multiplication (Mul) plus an addition (Add).


By contrast, if descending order is employed, as Eq.\,(2) in Fig.~\ref{fig:SU-FA}(a), the updating for $l_i^{(j)}$ merely requires one Exp and one Add. While such benefits may seem minor, the performance gain is substantial when large-scale parallel process long sequences. The procedure for the descending SU-FA is summarized in Fig.~\ref{fig:SU-FA}(b). Compared to the traditional FA and ascending SU-FA, the descending SU-FA on average reduces $25\%$ and $11\%$ complexity, respectively. In subsequent discussions, SU-FA defaults to adopt the descending order. Please note the inaccuracy of the predicted Max is co-optimized by the architecture in Section~\ref{subsection:SU-FlashAttention}.

\begin{algorithm}[t]\small
\label{alg:DSE}
\caption{\texttt{DSE for SOFA Tiling Size}.}
\textbf{Input:}~Evaluation function $\mathcal{L}$ and exploration space $\Theta$\;
\textbf{Initial:} Max Iter $T$, sample $\mathcal{D}_t=\{R_i,\mathcal{L}(R_i),i=1,...,n$\},
Best target function result $\mathcal{J}=\infty$\;
\While{$t<T$ \upshape and result does not converge}
{
$R_t\gets \mathop{\arg\max}\nolimits_{\Theta}\alpha (\Theta,\mathcal{D}_t)$,~$\mathcal{J}_{new}\gets\mathcal{L}(R_t)$\;
$\mathcal{D}_{t+1}\gets\{\mathcal{D}_t,(R_t,J_{new})\}$\;
$\mathcal{GP}_{new}\gets\texttt{Update}(\mathcal{GP},\mathcal{D}_{t+1})$\;
\If{$\mathcal{J}_{new}<\mathcal{J}$}{
$J\gets\texttt{Update}(J_{new})$\;
}
}
\end{algorithm}

\subsection{Design Space Exploration}\label{sec:DSE}
In the SOFA algorithm mechanism, the tiling size, i.e., $B_c$ in each layer and top-$k$ form an interesting design space. For larger $B_c$, i.e. smaller $T_c$ ($S=B_c\times T_c$), inference accuracy tends to increase. However, sorting complexity escalates significantly, yet the computation complexity of SU-FA decreases. We provide each of the hyperparameters with plenty of options as 1) $Tc_i$: $2-32$, step=2; 2) Top-$k$: $5\%-50\%$, step=$5\%$, to ensure that we can obtain a high-quality solution. However, such space is huge and unaffordable for brute force search. Taking BERT-Base with $12$ Transformer layers as an example, we need to search for the optimal choice in a $26$-dimensional space consisting over $10^{15}$ choices. Even though the inference on highly parallel GPU clusters costs less than $1$ ms, it will take unbearable time consumption (over $10^8$\,h) using traversal-based grid search for this remarkable design space. To this end, we apply a Bayesian optimization method to execute the search process. The targeted optimization problem (modeled as a Gaussian Process (GP) in Bayesian optimization) is constructed concerning both the accuracy and the computational complexity, which is formalized as Eq.\eqref{eq:Bayes}.
\begin{align}
\mathbf{minimize}~\mathcal{L}(R)&=\mathcal{L}_{en}+\alpha\times\mathcal{L}_{cmp}+\beta\times\mathcal{L}_{exp},\label{eq:Bayes}\\
\mathcal{L}_{cmp}&=\sum\nolimits_i(B_{ci}\cdot k)/\sum\nolimits_i(S\cdot k), \label{eq:Loss1}\\
\mathcal{L}_{exp}&=\sum\nolimits_i(S/B_{ci}),\label{eq:Loss2}
\end{align}

\noindent where $R$ is the hyperparameter vector composed of the $k$ and $B_{ci}$ of each layer, $\mathcal{L}_{en}$ is the cross-entropy loss, $\mathcal{L}_{cmp}$ and $\mathcal{L}_{exp}$ are the penalty terms for computation overhead, as defined in Eqs. \eqref{eq:Loss1} and \eqref{eq:Loss2}. $\alpha$ and $\beta$ are two coefficients to balance the accuracy and performance. The whole searching process is summarized in Alg.~\ref{alg:DSE}.

\begin{figure}[t]
\centering
\includegraphics[width=0.86\linewidth]{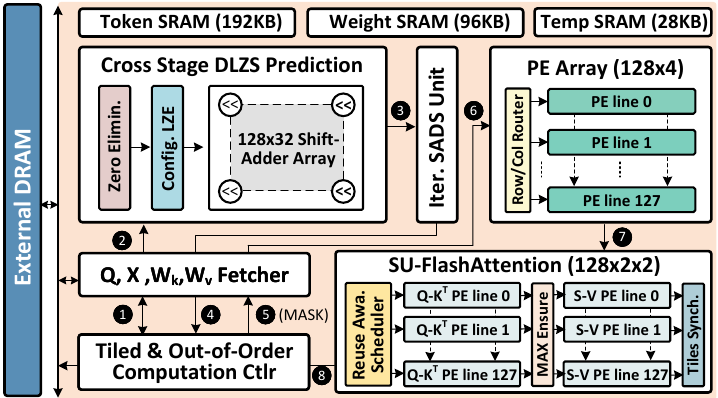}\vspace{-2mm}
\caption{High-level block diagram for the SOFA accelerator.}
\label{fig:SOFA_Hardware}\vspace{-4mm}
\end{figure}

\section{Architecture and Hardware Innovation}\label{sec:SOFA_Hardware}

Despite significant algorithmic acceleration, a naive implementation of SOFA encounters three challenges. First, LP is crucial in predicting vital tokens. It must ensure high precision and low power consumption. And the top-$k$ engine must support variable-length inputs and high throughput within low power overhead, due to the flexible tiling execution and high parallelism of LTPP. Second, specific architecture and datapath designs are needed to support the intra-stage operator-fusion paradigm of SU-FA for enhanced efficiency. Finally, during LTPP execution, the varying requirements of K and V for each query may lead to redundant memory access, thus necessitating a memory-efficient scheduling strategy. 
\subsection{Architecture Overview}\label{subsec:Architecture_overview}
Fig.~\ref{fig:SOFA_Hardware} depicts SOFA's overall architecture, which comprises six main modules: on-chip SRAM storage, a DLZS prediction unit, an iterative SADS unit, a PE array, an SU-FA unit, and a tiled $\&$ out-of-order controller. SOFA is designed to process $128$ queries in parallel. First, the indices of tokens and corresponding $\mathbf{W}_k$ of a tile produced by the \emph{controller} are sent to \emph{data fetcher}, which calculates the physical address and fetches data to on-chip SRAM \ding{202}. Then, the \emph{DLZS predictor} starts to estimate matrices $\hat{\mathbf{K}}$ and $\hat{\mathbf{A}}$ with log-based shift and summations \ding{203}. Next, the 128-row $\hat{\mathbf{A}}$ is sent to \emph{SADS unit}, to find out the top-$k$ important Q-K pairs \ding{204}. Subsequently, the sorting results are sent back to the \emph{controller} \ding{205}, which generates a top-$k$ mask, then \emph{data fetcher} reads corresponding data according to the mask \ding{206}. After that, the scheduler controls the \emph{PE array} to generate the necessary Ks and Vs \ding{207}. Later, the generated KVs are sent to the \emph{SU-FA unit} to execute compute-memory efficient attention calculations \ding{208}. Finally, the outputs of attention are stored to off-chip DRAM \ding{209}.               



\subsection{Reusable $\&$ Configurable DLZS Engine}\label{subsec:RDLZS}
\begin{figure}[t]
\centering
\includegraphics[width=0.99\linewidth]{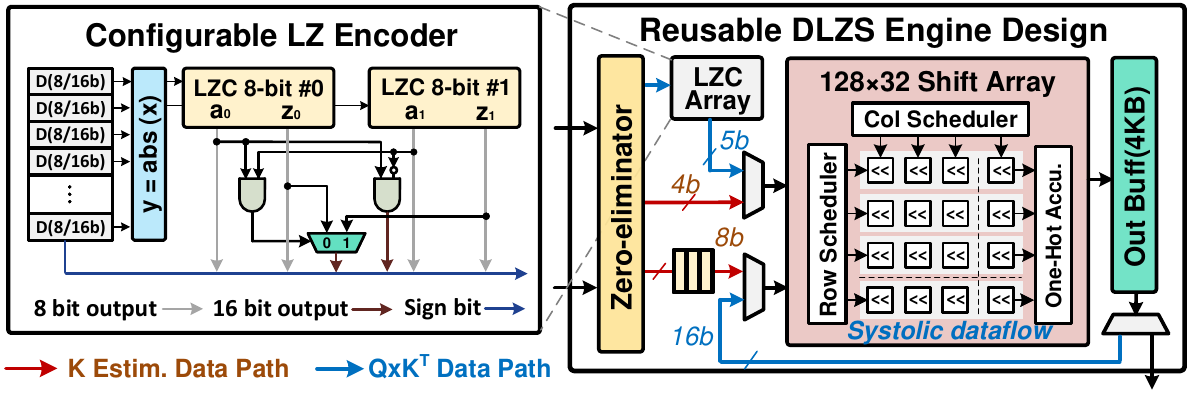}\vspace{-3mm}
\caption{Architecture for the cross-stage DLZS prediction.}
\label{fig:DLZS_Hardware}\vspace{-2mm}
\end{figure}

As discussed in Sec.~\ref{subsection:DLZS}, the DLZS unit is acquired to predict the $\mathbf{\hat{K}}$ and $\mathbf{\hat{A}}$, respectively. The two phases demand diverse precisions. In the former case, the inputs are $8$-bit token and weights, where the weights are pre-converted into LZ format. In contrast, the latter case requires operations with 16-bit precision, as the output of the former is truncated to at most $16$ bit. To this end, the LZE is designed as configurable to enable the 8$\&$16-bit mixed precisions. As depicted in Fig. \ref{fig:DLZS_Hardware} left, each LZE unit contains two 8-bit leading zero counters (LZCs)~\cite{milenkovic2015modular} connected in series. When the input is 8-bit, the two LZCs work independently. However, when the input becomes 16-bit, the two \emph{all-zero flag} $a_0$ and $a_1$ are performed through logic AND, then the corresponding output is employed as a selected signal to pick up 16-bit outputs. The processing flow of DLZS engine is illustrated in Fig. \ref{fig:DLZS_Hardware} right. First, the operands are sent to a zero eliminator module, where calculations with zeros are removed. Next, in the $\mathbf{K}$ prediction phase, 8-bit tokens and 4-bit LZ-format weights are transferred to the $128\times 32$ systolic shift array, and $\mathbf{\hat{K}}$ would be generated and cached in the output buffer. Then, in the $\mathbf{A}$ prediction phase, the 16-bit Qs are fed to the 16-bit mode LZC array. The generated 5-bit LZs along with the $\mathbf{\hat{K}}$ are sent to the shift array again, to produce the final estimated $\hat{\mathbf{A}}$.


\subsection{High-parallel and Flexible-input Supported SADS Engine}\label{subsection:Sorting}
As shown in Fig. \ref{fig:SOFA_workflow}, SOFA's tiled pipeline mechanism uses variable tile sizes for different models and tasks, requiring the sorting unit to handle flexible sub-segment lengths. This demands a sorting module that supports flexible inputs with low power overhead and high throughput to avoid bottlenecks. To this end, we design a flexible-input sorting architecture, with the high-parallel bitonic sorting core. Fig.~\ref{fig:SADS_Hardware} illustrates the SADS engine, which consists of two main modules:

\subsubsection{Sorting Module}
\begin{figure}[t]
\centering
\includegraphics[width=0.9\linewidth]{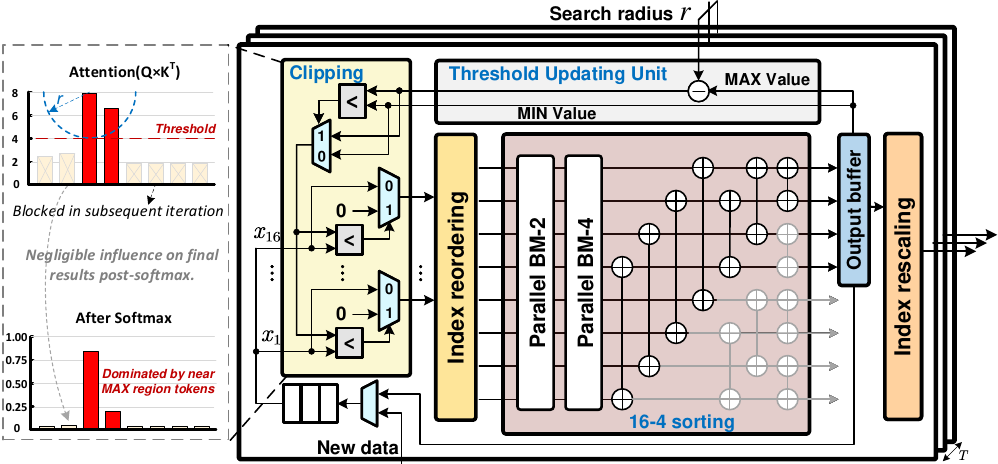}\vspace{-2mm}
\caption{Architecture for the flexible-input supported SADS engine.}
\label{fig:SADS_Hardware}\vspace{-4mm}
\end{figure}

The core sorting architecture uses a fully parallel $16$-to-$4$ bitonic sort design~\cite{lin2017hardware}. To handle flexible-length inputs, the module receives $12$ new inputs each time. combines them with the four largest values from the previous round, and outputs four new sorted values. After all elements are processed, the final results are generated. To reduce power consumption, we focus only on the top-k values and the top-1 and top-2 values, which are used to accelerate SU-FA. The order of the 3rd to k-th values is inconsequential, allowing us to eliminate redundant comparators, as shown in the shaded area in Fig.~\ref{fig:SADS_Hardware}.

\subsubsection{Clipping Module}
According to the proposed SADS in Sec.~\ref{subsec:SADS}, only elements in the feasible range are picked up and sorted accordingly. To this end, an adaptive clipping mechanism is implemented in this module to perform the filter function. As illustrated in Fig.~\ref{fig:SADS_Hardware}, it first reads the data to be sorted from \texttt{DLZS unit} and the threshold from \texttt{Threshold Updating (TU)}, respectively. The threshold is selected as the larger value between the \emph{top margin} ($=$Max-$r$) and the \emph{low bound} (The current Min value in the output buffer). In the beginning, both the \emph{low bound} and \emph{top margin} are set as zero and no values are eliminated. After obtaining the temporal sorted results, the \emph{low bound} and \emph{top margin} are updated in \texttt{TU} module. After that, the clipping mechanism is active and the smaller values are blocked in the following iterations. Given the efficiency of hardware implementation, we opt to directly substitute the blocked values with zeros. This approach effectively reduces power consumption from switching activities while maintaining hardware compatibility.

\subsection{Successive Updating FlashAttention Engine}\label{subsection:SU-FlashAttention}

\begin{figure}[t]
\centering
\includegraphics[width=0.8\linewidth]{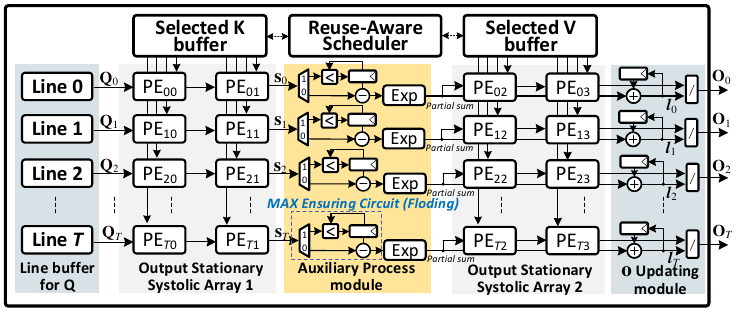}\vspace{-2mm}
\caption{The dedicated data flow architecture for the SU-FA mechanism.}
\label{fig:SU_FA_Hardware}\vspace{-4mm}
\end{figure}

While SU-FA can effectively reduce non-linear computations of traditional FA by leveraging the Max value provided by the \emph{top-$k$ stage}, it still faces a critical precision issue. This is because DLZS inherently is log-domain approximate computing, thus inevitably leading to estimation errors. Hence, hardware support is required to provide runtime assurance for the Max value. However, introducing a dedicated module for dynamic comparison directly would incur huge area overheads. To achieve this, we design a folded \texttt{auxiliary process} (AP) module capable of simultaneously supporting both Max value assurance functionality and synchronization between tiles (line 5-6 in Fig. \ref{fig:SU-FA}). As depicted in Fig. \ref{fig:SU_FA_Hardware}, this module operates in two configuration modes: \emph{computation} (0) and \emph{max update} (1). In \texttt{mode 0}, the intermediate value $s$ from the systolic array (SA) $1$ is directly subtracted with the Max value cached in Reg, and then fed to the Exp unit. Otherwise, in \texttt{mode 1}, the $s$ is sent into a comparator, compared with the Max cached in Reg, and the Reg's Max value is updated accordingly. Please note \texttt{Mode 1} is only activated during switching between different tiles or in the first computation phase within the same tile. The tiled computation controller manages the switching between the two modes.

\textbf{Workflow.} The SU-FA engine consists of four main parts: two SAs, an AP module, and an $\mathbf{O}$ updating module. First, the 128-row Q vectors are stored in the line buffer. Subsequently, two rows of K vectors corresponding to each Q vector are incrementally fed into SA-1, generating the corresponding $\mathbf{s}$. Then, $\mathbf{s}$ is sent into the AP module to perform the corresponding comparison or Exp calculation (Fig.~\ref{fig:SU-FA} line3,5,6), yielding intermediate partial sum results. The partial sum results are then fed into SA-2 and multiplied with the corresponding V vectors. Finally, the resulting output is sent to the $\mathbf{O}$ updating module to compute the final outputs (Fig.~\ref{fig:SU-FA} line 7).

\begin{figure}[t]
\centering
\includegraphics[width=0.92\linewidth]{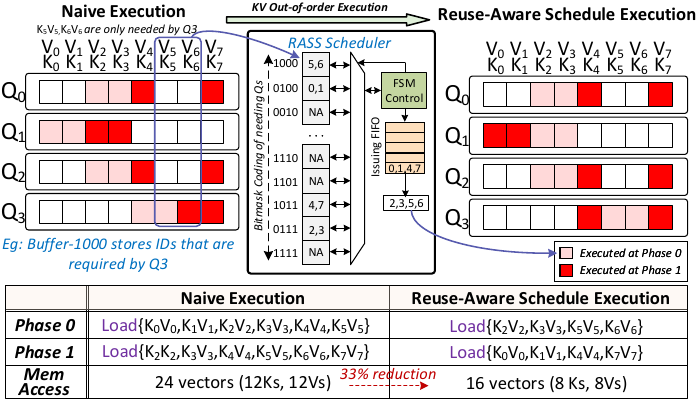}\vspace{-2mm}
\caption{Comparisons between RASS strategy and vanilla execution.}
\label{fig:locality_aware}\vspace{-4mm}
\end{figure}

\textbf{Reuse-Aware Schedule Scheme (RASS).} 
Due to dynamic sparsity, different queries select different Ks and Vs, with some overlap. Hence, how to effectively reuse K and V between different queries is a crucial challenge, especially in large-scale parallel processing. To this end, we propose \emph{reuse-aware schedule scheme (RASS)} with KV out-of-order execution to reduce overall memory access. As shown in Fig \ref{fig:locality_aware}, $k_2$ and $k_3$ are shared among three queries: $q_0$, $q_1$ and $q_2$, making them the top candidates for initial scheduling. Then, RASS seeks out Ks which are exclusively used by the remaining unscheduled query $q_3$, i.e., $k_5$ and $k_6$. As a result, $k_2$, $k_3$, $k_5$, and $k_6$ are packed together for execution in Phase 0. Such greedy search continues until all queries are allocated adequate Ks. As exemplified in Fig \ref{fig:locality_aware}, compared to the default left-to-right computation order, RASS reduces $33\%$ memory access.
  
We design a scheduler to implement the RASS. As shown in the middle of Fig.~\ref{fig:locality_aware}, the whole condition statement and control logic are implemented in an FSM controller. Besides, it involves a single-port read-write ID Buffer which is indexed using a bitmask of queries. For example, $k_5v_5$ and $k_6v_6$ are exclusively required by query $q_3$. Consequently, the pair '5,6' is stored in buffer-1000. Then the FSM controller accesses the ID Buffer according to the RASS, and dispatches the IDs into the Issuing FIFO in an optimized execution order.

\section{Evaluation}\label{sec:Evaluation}
\subsection{Experimental Setup}\label{subsec:Experimental_Setup}
We evaluate the soft performance of SOFA with several typical Transformer models and tasks by NVIDIA A100 GPU. For NLP tasks, the BERT-base and BERT-large models~\cite{devlin2018bert}, are selected and evaluated by eight tasks from GLUE~\cite{wang2018glue} and SQuAD v1.1~\cite{rajpurkar2016squad}. The maximum sequence length for BERT-B/L is 256/256/384/512/512 for MRPC/RTE/SQUAD/STSB/QNLI, respectively. Moreover, for GPT-2~\cite{radford2019language}, Bloom-1.7B~\cite{le2022bloom}, Llama7B/13B, language modeling tasks on Wikitext-2\cite{merity2016pointer}, WikiLingua\cite{ladhak2020wikilingua}, Wiki-raw and Winogrande\cite{sakaguchi2021winogrande} are evaluated. The maximum length for datasets on evaluated Bloom1.3B/Llama7B/13B is 2k/4k/4k, respectively. For CV tasks, we choose the latest PVT (with 3192 sequence length)\cite{wang2021pyramid} for ImageNet-1k classification~\cite{deng2009imagenet} by fine-tuning the checkpoint of ImageNet-21k. All models are implemented with Pytorch libraries~\cite{paszke2017automatic} and Huggingface Transformer project~\cite{wolf2020transformers}. For each task, we execute fine-tuning on NVIDIA A100 GPU after token pruning to recover accuracy. 


For hardware evaluation, we performed the RTL design for the  SOFA accelerator and utilized Synopsys DC on TSMC $28$nm CMOS technology, to estimate the logic parts'area and power consumption. The power, area, and read/write bandwidth of on-chip SRAM buffers are estimated through CACTI\cite{muralimanohar2009cacti}. For modeling off-chip DRAM, we utilize Ramulator\cite{kim2015ramulator} to simulate the memory behaviors and employ the same method with~\cite{cavigelli2016origami,andri2016yodann,wang2017energy} to estimate the IO power. According to the synthesized results, the latency of the critical path is less than $1$ ns. Then, we assume the running frequency of SOFA is $1$ GHz. We extract each stage’s actual cycles by simulating the RTL with Verilator~\cite{snyder2004verilator}, based on which a cycle-level simulator is implemented to evaluate end-to-end performance.

For comparisons with GPU, we deploy the benchmarks on the A100 platform using the Pytorch framework. We measure execution time by inserting \texttt{torch.cuda.synchronize} at the start and end points, and then calculate the elapsed time. For power measurement, based on \texttt{nvidia-smi}, we first measure the system’s idle power, and then repeatedly run workloads and get the total power. The dynamic power is total power minus idle power. Based on the computational workload, we derive the average throughput and energy efficiency. Similarly, we run the cloud TPU\cite{jouppi2017datacenter,google_cloud_tpu} to analyze the performance on diverse commercial hardware.



\subsection{Algorithm Performance}\label{subsec:Alg_Evalu}

\begin{figure}[t]
\centering
\includegraphics[width=0.65\linewidth]{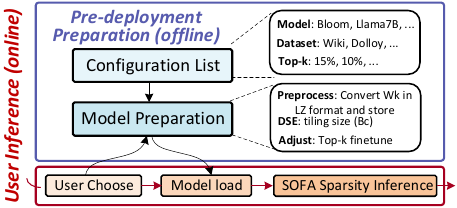}\vspace{-3mm}
\caption{The preparation and execution flow diagram of SOFA.}
\label{fig:flow-diagram}\vspace{-1mm}
\end{figure}

Fig.\ref{fig:flow-diagram} illustrates the SOFA flow-diagram, comprising two phase: \emph{Pre-deployment Preparation (PP)} and \emph{User Inference (UI)}. In the PP phase, the server selects models and corresponding datasets, then preprocesses each model through DSE and fine-tuning. All processed models are stored for user selection. In UI phase, users simply select their desired model, which, once loaded, enables real-time dynamic sparsity inference with SOFA.

\subsubsection{Algorithm Settings} 
In DSE objective function\eqref{eq:Bayes}, the coefficient $\alpha$ adjusts the proportion of the increased sorting cost, while $\beta$ controls the proportion of the benefit from reduced exponential operations. Initially, we conducted numerous experiments on BERT/PVT/GPT-2/Bloom/LLama to determine the search range for each hyperparameter. Subsequently, during training, we employed grid search to find the optimal parameter for each model and applied the successive halving method to accelerate the process.
According to our experiments, the $\alpha$/$\beta$ is set as 0.24/0.31 (BERT-B/L), 0.2/0.24 (ViT), 0.4/0.42 (GPT-2), 0.53/0.56 (Bloom-1.7B), and 0.58/0.63 (LLama-7B/13B), respectively. We then search for $200$ iterations with each learning rate (1e-1,5e-2,1e-3) to obtain the optimal tiling setting.

\subsubsection{Overall Performance} 
\begin{figure}[t]
\centering
\includegraphics[width=0.999\linewidth]{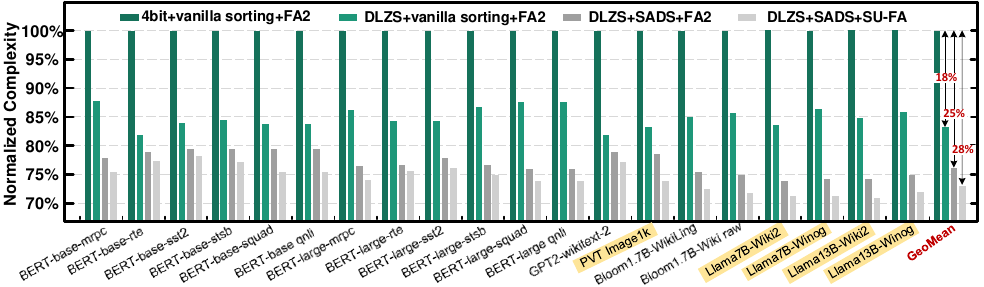}\vspace{-3mm}
\caption{Complexity reduction for the proposed DLZS, SADS and SU-FA.}
\label{fig:complexity_reduction}\vspace{-2mm}
\end{figure}

We first set an ablation experiment to evaluate the low-complexity advantages of DLZS, SADS and SU-FA by comparing them with a baseline scheme. The baseline is assumed to utilize 4-bit multiplications in \emph{pre-compute stage}, vanilla sorting in \emph{top-k stage} and traditional FA in \emph{formal-compute stage}. The complexity for different operations is normalized by the arithmetic complexity model~\cite{brent2010modern}. For fairness, each model's loss remains under $2\%$. As shown in Fig.\ref{fig:complexity_reduction}, DLZS reduces complexity by $18\%$ on average compared to the baseline. The reduction mainly comes from its multiplier-free computing and half-conversion feature. Further, SADS and SU-FA contribute to an extra $10\%$ reduction through segmented sorting and simplifying redundant non-linear computations using top-$k$ information. Overall, compared to traditional mechanisms, SOFA's software strategy achieves $28\%$ lower computation complexity under the same token sparsity, making SOFA accelerator effective for handling the LTPP scenario.

\begin{figure*}[t]
\centering
\includegraphics[width=0.93\linewidth]{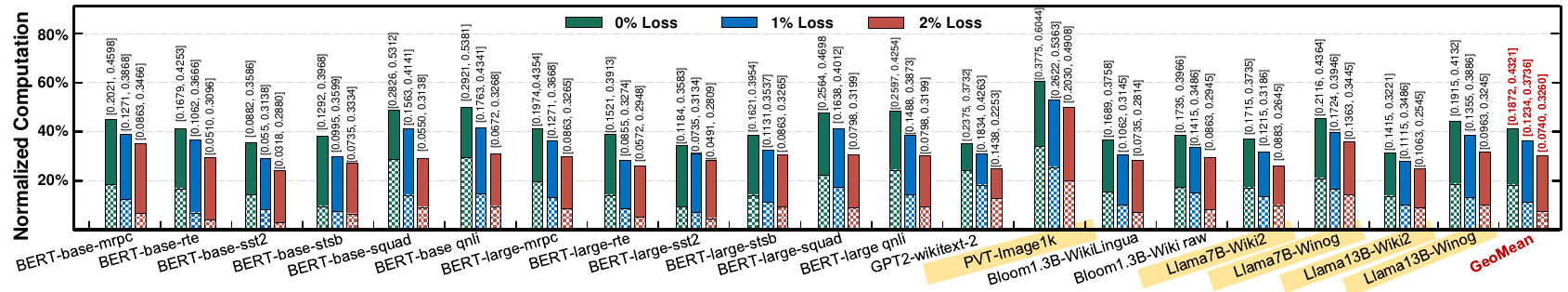}\vspace{-4mm}
\caption{Computation reduction by LP with diverse loss tolerance. [X, Y] respectively denote the computation reduction for the Atten part and QKV+Atten.}
\label{fig:pruning_reduction}
\end{figure*}

To demonstrate the effectiveness of SOFA in detecting token sparsity, Fig.\ref{fig:pruning_reduction} shows the QKV and attention computation reduction, introduced by the SOFA's sparsity prediction (LP). For practicality, we statistically analyzed the reduction in computational workload while ensuring accuracy losses remained below $0\%$, $1\%$, and $2\%$ respectively. Different end-to-end metrics are utilized for evaluation, such as F1 score for SQuAD and accuracy for RTE. 
On average, SOFA's sparsity prediction can reduce the attention+QKV computation by $56.8\%$/$62.6\%$/$67.4\%$ with $0\%$/$1\%$/$2\%$ accuracy loss, respectively. Focusing solely on the attention part, SOFA reduces computation by $81.3\%$/$87.7\%$/$92.6\%$.

\textbf{Discussion on accuracy:}
In Top-K pruning, there is a hyperparameter $k$. Lowering $k$ eliminates more QK-pairs, which in turn reduces computation. However, reducing $k$ too aggressively could lead to the exclusion of some relatively important QK-pairs, thus hurting the model's accuracy. Moreover, different datasets exhibit varying features of sparsity due to their distinct data types and tasks. Consequently, in the pre-deployment preparation, the value of $k$ can be modified to optimize the algorithm's exploiting of sparsity to minimize computation while maintaining accuracy. For example, we observed that datasets like SST2 and STS-B, used for sentiment classification or semantic analysis, typically exhibit high sparsity because one or two keywords often indicate sentiment. Therefore, their computation reduction is adjusted to 90\% while the accuracy loss is controlled within 1\%. In contrast, image datasets generally contain a high amount of key information and have lower data redundancy compared to text classification datasets, resulting in lower sparsity. As a result, their computation reduction is adjusted to 73\% with a 1\% accuracy loss.


\subsection{Architecture Evaluation}\label{subsec:Arch_Evalu}

\begin{figure}[t]
\centering
\includegraphics[width=0.999\linewidth]{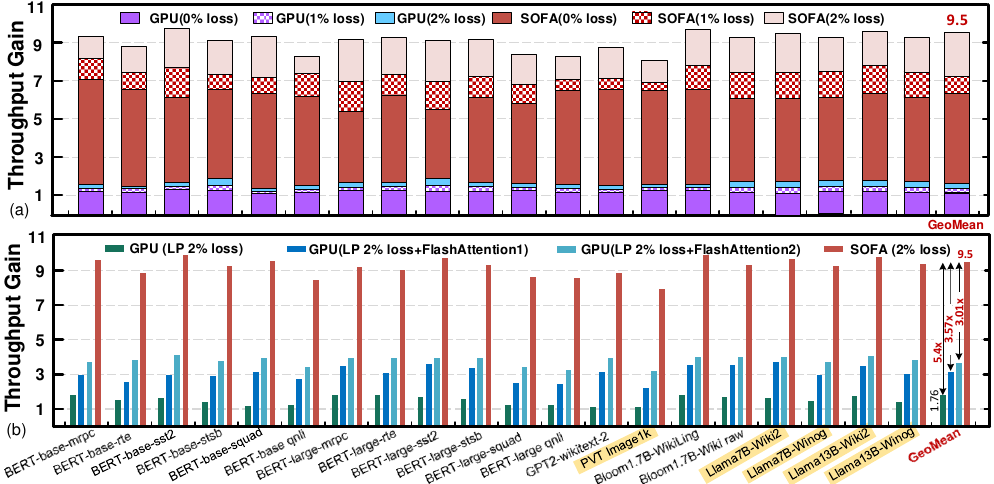}\vspace{-2mm}
\caption{Throughput gain of SOFA over (a) LP (b) LP+FA-1/2 on A100 GPU.}
\label{fig:throughput_improve}\vspace{-4mm}
\end{figure}

\begin{figure}[t]
\centering
\includegraphics[width=0.999\linewidth]{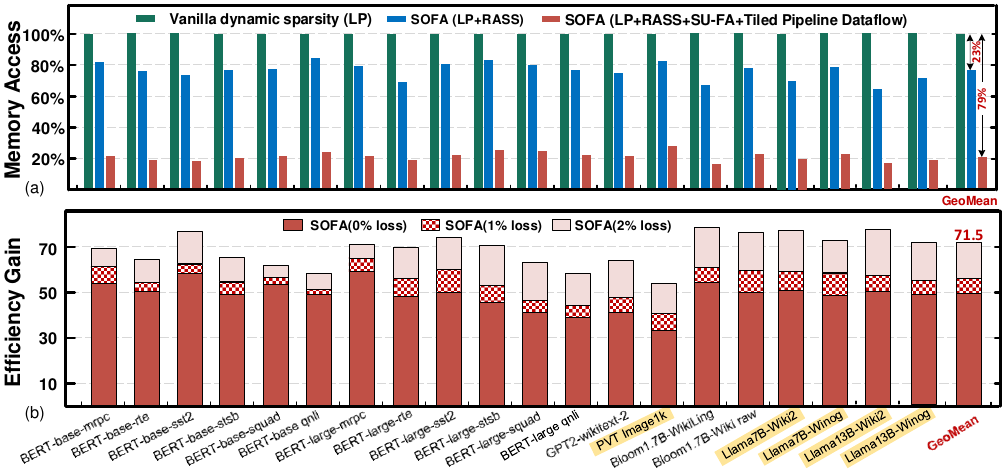}\vspace{-2mm}
\caption{(a) Memory access reduction of SOFA. (b) Efficiency gain of SOFA over Nvidia A100 GPU.}
\label{fig:efficiency_improve}\vspace{-2mm}
\end{figure}

\emph{Throughput Improvement:} Fig.~\ref{fig:throughput_improve} (a) compares the throughput of SOFA with A100 GPU on all benchmarks versus diverse accuracy loss. As can be seen, LP enables $1.08$-$1.78\times$ of speed up on GPU with its sparsity detection. Unfortunately, the GPU cannot leverage the LP results as it cannot handle high sparsity or fine-grained on-demand KV calculations. Nor can it run the cross-stage DLZS-based prediction efficiently. By contrast, the SOFA exhibits an average $85.2\%$ PE utilization due to its stage-fused fine-grained tiled dataflow, which pipelines cross-stage DLZS prediction, SADS sorting, and SU-FA, leading to almost triple sparsity utilization than GPU. Further, the SU-FA engine is tailored to support sparsity attention acceleration with reduced computational complexity. Overall, SOFA achieves $6.1\times$, $7.2\times$ and $9.5\times$ inference speed up with $0\%/1\%/2\%$ accuracy degradation. Fig.\ref{fig:throughput_improve} (b) further compares the SOFA with LP+traditional FA and LP+FA2 on A100. On average, FA on GPU brings around $1.5\times$ gain, leading to a total $2.7\times$ speed up combined with LP. By adjusting the loop order to avoid some factor scaling nonlinear computations, FA2 achieved a further $1.19\times$ throughput improvement. However, due to the difficulty of fine-grained cross-stage data movement on GPUs and the challenges of optimizing FA1/2 to support fine-grained scheduling and sparse computation, it is difficult to achieve higher improvements. By contrast, SOFA (soft+archi) achieves $9.5\times$ gain, which is $3.01\times$ greater than vanilla LP+FA2 on GPU. Fig.\ref{fig:efficiency_improve} (a) shows the memory redcution effectiveness of SOFA. Compared with the baseline with vanilla dynamic sparsity, SOFA with RASS can reduce average $23\%$ memory access. With SU-FA and tiled dataflow, the reduction rises further $79\%$.  

\begin{figure}[t]
\centering
\includegraphics[width=0.97\linewidth]{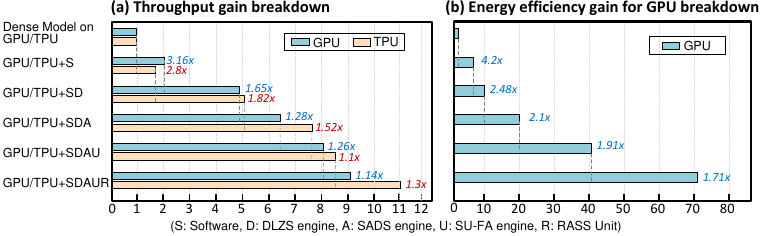}\vspace{-2mm}
\caption{(a) Throughput gain of GPU/TPU with SOFA's mechanisim (b) Energy efficiency gain of GPU with SOFA's mechanisim.}
\label{fig:accleration_breakdown2}\vspace{-2mm}
\end{figure}

Fig.~\ref{fig:accleration_breakdown2} (a) illustrates the breakdown of throughput improvement achieved by GPU A100 and TPU with the hardware-software mechanism of SOFA. The baseline is executing a dense Transformer model on GPU/TPU. With SOFA software optimization, GPU and TPU achieve improvements of $3.16\times$ and $2.9\times$, respectively. Both of them cannot fully leverage all the benefits of SOFA software. GPU performances better than TPU due to its ability to better handle some of the fine-grained computations and scheduling in SOFA software. Adding SOFA's engines incrementally, we observed significant performance gains. The GPU with the DLZS engine achieves a $1.6\times$ speedup due to the systolic data flow improving data reuse, which the GPU's vector engine cannot support. The TPU with the DLZS engine shows an even higher improvement of $1.82\times$ because its limited control instructions are inefficient at handling DLZS's logical branching. Similarly, the SADS engine, with its customized data paths, achieves a $1.28\times$ improvement on the GPU and $1.52\times$ on the TPU by quickly and efficiently executing redundant computations. Further, the SU-FA engine improves performance by $1.26\times$ on the GPU and $1.1\times$ on the TPU due to its max-assured circuits that avoid inefficient recomputation and data movement caused by log-domain calculation errors. The SU-FA engine employs a systolic array design. Since the TPU's support for systolic arrays is inferior to the GPU's, it gains a greater speedup than GPU. Lastly, the RASS unit achieves improvements of $1.14\times$ on the GPU and $1.3\times$ on the TPU due to its customized control unit enabling more efficient scheduling and data arrangement.

\begin{table*}[ht]
\renewcommand{\arraystretch}{0.93}
    \centering
    \scriptsize
    \caption{Summary and Comparison with SOTA works.}\vspace{-3mm}
    \begin{tabular}{l||ccc|cccccccccc}
        \hline
        \multirow{3}{*}{\!\!\textbf{Accelerators}}  & \multicolumn{3}{c|}{\multirow{1}{*}{\textbf{Software Performance}}} & \multicolumn{6}{c}{\multirow{1}{*}{\textbf{Hardware Performance}}} \\ 
        \cline{2-14}
        %
        
        & \multicolumn{1}{c|}{\multirow{2}{*}{Sparsity$^{\text{\ding{172}}}$}} & \multicolumn{1}{c|}{\multirow{1}{*}{Accu}} & \multicolumn{1}{c|}{\multirow{1}{*}{Saved}} & \multicolumn{1}{c|}{\multirow{1}{*}{Tech}} & \multicolumn{1}{c|}{\multirow{1}{*}{Freq}} & \multicolumn{1}{c|}{\multirow{1}{*}{Area}} & \multicolumn{2}{c|}{\multirow{1}{*}{Power [W]}}  & \multicolumn{1}{c|}{\multirow{1}{*}{Throup.}}  & \multicolumn{2}{c|}{\multirow{1}{*}{Energy Effi.$^{\text{\ding{174}}}$ [GOPS/W]}} & \multicolumn{1}{c|}{\multirow{1}{*}{Area Effi.$^{\text{\ding{174}}}$}} & \multicolumn{1}{c}{\multirow{1}{*}{Latency}} \\
        \cline{8-9}
        \cline{11-12}
        
        & \multicolumn{1}{c|}{} &  \multicolumn{1}{c|}{Loss} & \multicolumn{1}{c|}{Comp$^{\text{\ding{173}}}$} & \multicolumn{1}{c|}{[nm]}  & \multicolumn{1}{c|}{[Hz]} & \multicolumn{1}{c|}{[mm$^2$]} & \multicolumn{1}{c|}{Core}  & \multicolumn{1}{c|}{IO} & \multicolumn{1}{c|}{[GOPS]} & \multicolumn{1}{c|}{Core} & \multicolumn{1}{c|}{Device$^\dagger$} & \multicolumn{1}{c|}{[GOPS/mm$^2$]} & \multicolumn{1}{c}{[ms]} \\

        \hline
        \rowcolor{mygray}\!\!\! $\mathbf{A}^3$\!\cite{ham20203}    & \multicolumn{1}{c|}{Unstr}  & $5.3\%$  & 40$\%$ & 40 & 1G & 2.08 & 0.205 & 0.617 & 221 & 1863 & \textbf{300}
        & \textbf{217} & \textbf{622} \\  \hline
        
        \!\!\!\textbf{ELSA}\cite{ham2021elsa} &   \multicolumn{1}{c|}{Unstr} &  $2\%$ & 73$\%$ & 40 & 1G & 1.26 & 0.969 & 0.525 & 1090 & 1944 & \textbf{1004} & \textbf{1765} & \textbf{252} \\ 
        \hline
        
        \rowcolor{mygray}\!\!\! \textbf{Sanger}\cite{lu2021sanger}   & \multicolumn{1}{c|}{Str} & $0\%$ & 76$\%$ & 55 & 500M & 16.9 & 2.76 & - & 2285 & 2342 & - & \textbf{522} & \textbf{241} \\  
        
        \hline
        
        \!\!\!\textbf{DOTA}\cite{qu2022dota}  & \multicolumn{1}{c|}{Str} & $0.8\%$ & 80$\%$ & 22 & 1G & 4.44 & 3.02 & -
        & 4905 & 817 & - & \textbf{683} & \textbf{448} \\  \hline
        
        \rowcolor{mygray}\!\!\!\textbf{Energon}\cite{zhou2022energon} &        \multicolumn{1}{c|}{Unstr} &  $0.9\%$ & 77$\%$ & 45 & 1G & 4.2 & 0.32 & 2.4 & 1153 & 7007 & \textbf{450} & \textbf{709} & \textbf{477}\\
         
         \hline
         \!\!\!\textbf{DTATrans}\cite{yang2022dtatrans} & \multicolumn{1}{c|}{Unstr} &  $0.74\%$ & 74$\%$ & 40 & 1G & 1.49 & 0.734 & - & 1304 & 3071 & - & \textbf{1786} & \textbf{652} \\  \hline
         
         \rowcolor{mygray}\!\!\!\textbf{SpAtten}\cite{wang2021spatten}   & \multicolumn{1}{c|}{Str} &  $0.9\%$ & 67$\%$ & 40 & 1G & 1.55 & 0.325 & 0.617 & 360 &  1915 & \textbf{447} & \textbf{474} & \textbf{382} \\  \hline

         \!\!\!\textbf{FACT}\cite{qin2023fact}  & \multicolumn{1}{c|}{Unstr}  &  $0\%$ & 79$\%$ &  28 & 500M & 6.03 & 0.337 & - & 928 & 2754 & - & \textbf{154} & \textbf{296} \\  \hline

         \rowcolor{mygray}\!\!\!\textbf{SOFA}  &    \multicolumn{1}{c|}{Unstr} & $0\%$ & 82$\%$ &  28 & 1G & 5.69 & 0.95 & 2.45 & 24423 & 25708 & \textbf{7183} & \textbf{4292} & \textbf{45} \\  \hline
    \end{tabular}
    \begin{tablenotes}
    \vspace{-0.7mm}\scriptsize
    \item \!\!\! $^{\text{\ding{172}}}$ Unstructured or Structured sparsity. ~~ $^{\text{\ding{173}}}$ Comp saving = Reduced attention computaion - Prediction computation. ~~$^\dagger$ Device= IO+Core.
    \item \!\!\! $^{\text{\ding{174}}}$ Scaled to 28nm and 1.0V CMOS with $f\propto 1/s^2$ and power (core) $\propto (1/s)(1.0/Vdd)^2$, where s=Tech/28nm\cite{wang2017energy,liu2020energy}
    \end{tablenotes}
    \label{tab:hardware_norm}\vspace{-2mm}
\end{table*}

\emph{Area, Power and Energy:} Table~\ref{tab:core_hardware_result} shows the power and area breakdown of SOFA accelerator. It has a total area of $5.69$ mm$^2$, and LP accounts for merely $18\%$ and $15\%$ of area and power. This benefits from the multiplier and converter-free design in DLZS engine and the low-overhead design of SADS engine. Fig.~\ref{fig:efficiency_improve}(b) illustrates the overall energy-efficiency gain of SOFA compared to the A100 GPU. On average, SOFA achieves $49.8\times$, $57.6\times$, and $71.5\times$ greater energy efficiency in comparison to the A100 GPU with $0\%$, $1\%$ and $2\%$ accuracy loss, respectively. In Fig.~\ref{fig:accleration_breakdown2} (b), we also show the efficiency gain breakdown. DLZS and SADS engines bring $2.48\times$ and $2.1\times$ efficiency gain, respectively. Further, SU-FA and RASS units together bring about $3.27\times$ gain. In Table~\ref{tab:DRAM_hardware_result}, we list the power overhead consumed by the memory interface~\cite{leibowitz20104} and external DRAM.

\begin{table}[t]
\renewcommand{\arraystretch}{0.9}
\caption{Area and Power breakdown for SOFA (Core Part) at $1$GHZ.}\vspace{-4mm}
\begin{center}
\begin{tabular}{l|ccc}
\specialrule{0.12em}{0.5pt}{1pt}
\!\!\textbf{Modules} & \textbf{Parameters} & \!\!\textbf{Area[mm$^2$]}\!\! & \!\!\!\textbf{Power[mW]} \\
\hline
\!\!\multirow{2}{*}{DLZS prediction}\!\! & \!\!$128$$\times$$32$ shift PEs\!\! & \multirow{2}{*}{$0.351$} & \multirow{2}{*}{$29.05$}  \\
 & \!\!$128$ LZEs \!\! &  &   \\
 \hline
\!\!\multirow{2}{*}{Iterative SADS} & \!\!\!$128$ $16$-$4$ sort cores& \multirow{2}{*}{$0.679$} & \multirow{2}{*}{$112.79$} \\
\!\!& \!$128$ clipping units\! &  &   \\
\hline
\!\!KV generation & \!$128$$\times$$4$ $16$\,bit PEs\! & $0.875$ & $146.21$ \\
\hline
\!\!\multirow{3}{*}{SU-FA module} & \!$128$$\times$$4$ $16$\,bit PEs\! & \multirow{3}{*}{$3.012$} & \multirow{3}{*}{$485.12$}  \\
\!\! & $128$ EXP units\! &  &   \\
& $128$ DIV units &   &    \\
\hline
\!\!\multirow{3}{*}{Memory}  & \!\!\!$192$KB\,Token\,SRAM\!\! & \multirow{3}{*}{$0.497$} & \multirow{3}{*}{$170.23$}  \\
\!\! & \!\!$96$KB\,Weight\,SRAM\!\! &   &    \\
\!\! & \!\!$28$KB\,Temp\,SRAM\!\! &   &    \\
\hline
\!\!Scheduler $\&$ Others & - &  $0.280$ & $6.45$  \\
\hline
Off-Chip DRAM & \multicolumn{3}{c}{HBM2, 16$\times$ HBM channels $@$ 2GHz} \\
\hline
\textbf{Total} &  \multicolumn{3}{c}{\!\!TSMC $28$nm: Area=$5.69$mm$^2$, Power=$949.85$mW} \\
\specialrule{0.12em}{0.1pt}{0.1pt}
\end{tabular}
\end{center}
\label{tab:core_hardware_result}
\vspace{-4mm}
\end{table}

\begin{table}[t]
\renewcommand{\arraystretch}{0.9}
\caption{Power breakdown of SOFA.}\vspace{-4mm}
\begin{center}
\scriptsize
\begin{tabular}{l|m{1.4cm}<{\centering}m{2.4cm}<{\centering}m{1.0cm}<{\centering}m{1.0cm}<{\centering}}
\specialrule{0.12em}{0.5pt}{1pt}
 & \textbf{Core Part} & \textbf{Memory Interface} & \textbf{DRAM} & \textbf{Overall}\\
 \hline
\textbf{Power} & $0.95$W & $0.53$W & $1.92$W & $3.40$W \\
\specialrule{0.12em}{0.5pt}{0.5pt}
\end{tabular}
\begin{tablenotes}
    \vspace{-0.7mm}\scriptsize
    \item \!\!\! $^{\text{\ding{172}}}$ {The DRAM and Interface power are estimated with $59.8$GB/s.} 
    \end{tablenotes}
\end{center}
\label{tab:DRAM_hardware_result}\vspace{-2mm}
\end{table}

\subsection{Comparison with Existing Acclerators}
FACT, Sanger, Energon, SpAtten, ELSA and DOTA are SOTA Transformer dynamic sparsity accelerators. However, their designs focus on computational optimization, overlooking that memory access becomes the de facto bottleneck after computational optimization. The head pruning technique in SpAtten can partly alleviate memory access issues, but its efficiency is limited as it fundamentally depends on the characteristics of the task. On the other hand, although Energon considers a certain computation-to-memory access ratio in its architecture design, it still suffers from inefficiency due to the variability of computational tasks and models. In summary, previous efforts still lack simultaneously optimizing both computation and memory access. When imbalance arises between computation and memory access due to sparsity, it hampers further enhancement of hardware efficiency. SOFA employs a holistic FlashAttention-like scheme to divide all work stages of dynamic sparsity into fine-grained tile manner, and leverages the sort information for cross-stage collaborative optimization. 
Table~\ref{tab:hardware_norm} compares the features of software and hardware performance across these SOTA accelerators. Benefiting from the low complexity of LP mechanism, SOFA achieves the greatest reduction (82\%) in computation at the same accuracy loss of 0\%. We list their hardware parameters and present a normalized comparison\cite{wang2017energy,liu2020energy} of energy and area efficiency. Compared with these SOTA accelerators, the device(core+IO) energy efficiency of SOFA is $7183$ GOPS/W, marking a substantial improvement of $7.2\times$ to $24\times$. This improvement stems from the fine-grained data flow achieved through collaborative cross-stage optimization, which effectively reduces off-chip memory accesses. Additionally, SOFA achieves $4292$ GOPS/mm$^2$ area efficiency, which is $2.4\times$ to $27.9\times$ better than the SOTA accelerators. The gain in area efficiency primarily arises from the algorithm-hardware co-optimization for low complexity. 


We also quantitatively compare the latency of the SOTA accelerators by evaluating them to execute an attention part (137GOPs) of Llma7B. For fairness, all accelerators are scaled to 128 multipliers clocked at 1GHz. For example, the effective throughput of FACT is 928 GOPS in 500MHz with 512 multipliers. Then its execution latency is 2*137/928=296ms. Compared to the $0\%$ loss accelerators FACT and Sanger, SOFA achieves $6.6\times$ and $5,4\times$ latency reduction, respectively. Moreover, SOFA achieves $8.5\times$ and $10.6\times$ latency decrease over SpAtten and Energon, respectively. Such reduction in SOFA latency is mainly attributed to the fine-grained tiling execution across stages, as shown in Fig.\ref{fig:SOFA_workflow}.


\section{Related Works and Discussion}
\textbf{Efficient Transformer Accelerator.} Numerous studies~\cite{fang2022algorithm,ham20203,ham2021elsa,hong2022dfx,li2022accelerating,lu2021sanger,wang2021spatten,qu2022dota,yang2022dtatrans,zhou2022energon,yazdanbakhsh2022sparse,zadeh2020gobo,qin2023fact,li2020ftrans} have been proposed to improve the energy efficiency and speed of Transformer inference. However, most of these works focus on attention computation reduction, including static sparsity~\cite{you2023vitcod,li2020ftrans,shen2022salo,fan2022adaptable}, dynamic sparsity~\cite{ham20203,ham2021elsa,lu2021sanger,wang2021spatten,qu2022dota,yang2022dtatrans,zhou2022energon,qin2023fact} and hybrid sparsity~\cite{zhao2024hardware}. However, when computation is optimized, the memory access would dominate the overall power and time, especially for LTPP scenarios, which these works ignore. By contrast, SOFA optimizes both compute and memory access, thus greatly outperforming previous works. Further, all dynamic sparsity efforts focus on individually optimizing each stage for higher efficiency. Unlike these works, SOFA exhibits a cross-stage holistic optimization. This provides SOFA with an ever-overlooked opportunity for cross-stage tiling, executing a fine-grained tiled dataflow that accelerates inference while reducing off-chip memory access.

\textbf{Neural network accelerator with sparsity.} There are very many ASIC or FPGA accelerators~\cite{hojabr2021spaghetti,gondimalla2019sparten,asgari2020alrescha,deng2021gospa,qin2020sigma,gudaparthi2022candles,fang2022algorithm,hanson2022cascading,lew2022anticipating,li2022ristretto,li2021escalate,li2022accelerating,liu2022s2ta,pavon2021via,rucker2021capstan,sadi2019efficient,walia2021fast,yang2020procrustes,Gon2023Eureka} that leverage sparsity to optimize the performance of neural network inference. There also exist general sparse tensor algebra accelerators~\cite{srivastava2020matraptor,mahmoud2020tensordash,kwon2019tensordimm,hegde2019extensor,kanellopoulos2019smash,chen2020tpspmv} proposed in recent years, which can be used to process sparse FC layers. Recently, works~\cite{wu2022sparseloop,wu2023highlight,shin2022griffin} utilize hierarchical sparsity to construct a comprehensive design space and provide accurate performance metrics, which enable the automatic and optimal design of sparse DNN accelerators. However, most of the works focus on exploiting pre-trained static sparse weights. By contrast, SOFA leverages LP to predict on-the-fly dynamic sparsity. Especially, such sparsity comes from the \emph{argmax approximation} property of softmax, thus needing to be detected actively. This makes the traditional near-zero-based sparsity methods inapplicable. Through recently some works are config for activation sparsity \cite{jang2021sparsity} and both weight and activation sparsity \cite{wu2023highlight,huang2023rm,wang2021dual}, they are all based on the near-zero sparsity, thus failing to the top-$k$ sparsity scene, which is SOFA targets.  

\textbf{Fused operator tiling accelerators.}  Many works~\cite{alwani2016fused,goetschalckx2021depfin,im2020dt,lee2019full,min2021dadu,mo202112,zhang2018dnnbuilder,mei2023defines} leverage layer-fusion strategy to optimize the DNN inference performance. 
Specifically, DNNBuilder\cite{zhang2018dnnbuilder} and DeFiNES\cite{mei2023defines} use a depth-first-like layer fusion in CNNs to enhance data reuse via cross-layer tiling, enabled by the weak operator dependencies in CNNs. However, \emph{dynamic sparsity} of Transformers face bottlenecks due to row dependency in the top-$k$/softmax operator, restricting dynamic sparsity for long sequences. SOFA addresses this by employing the DCE data distribution property, unlocking the possibility of depth-first-like execution in Transformer \emph{dynamic sparsity} for the first time. DeepBurning~\cite{cai2022deepburning} partitions NN graphs at the inter-operator granularity and executes them in a pipeline fashion. In contrast, SOFA achieves finer-grained execution by dividing within the operator, leading to more efficient SRAM utilization. FLAT \cite{kao2023flat} fuses the two matmul operators and softmax in attention to reduce off-chip memory access but fails to resolve softmax row dependency. Traditional FlashAttention \cite{dao2022flashattention,dao2023flashattention} successfully unlocks the row dependency of softmax but at the expense of surging computation costs. In this aspect, SOFA leverages SU-FA to successfully solve the row dependency in softmax, allowing for finer-grained tiling and reducing SU-FA complexity using top-$k$ sorting information.



\section{Conclusion}\label{sec:Conclusion}
We propose SOFA, a cross-stage compute-memory efficient algorithm-hardware co-design to accelerate dynamic sparsity Transformer inference for LTPP. We introduce a novel log-domain DLZS computing paradigm to estimate Q-K pairs with add-only operation, requiring less converters. To prevent memory access from becoming a bottleneck after sparsity computation optimization, we propose SADS and SU-FA to enable cross-stage tiling for the end-to-end workflow. Leveraging this tiling strategy, SOFA executes a fine-grained pipeline dataflow across diverse stages, effectively mitigating memory access and latency issues. Efficient architecture is designed to support and accelerate the above mechanisms with a memory-efficient reuse-aware schedule. SOFA achieves $71.5\times$ energy saving than Nvidia A100 GPU, and $15.8\times$ higher energy efficiency than 8 SOTA accelerators, respectively.

\bibliographystyle{IEEEtranS}
\bibliography{Micro-SOFA}

\end{document}